\documentclass[preprint,amssymb,nofootinbib,amsmath,11pt]{JHEP3}

\JHEPspecialurl{http://jhep.sissa.it/JOURNAL/JHEP3.tar.gz}

\usepackage{epsfig,multicol,amsmath}

\def\beq{\begin{equation}}
\def\eeq{\end{equation}}
\def\bea{\begin{eqnarray}}
\def\eea{\end{eqnarray}}

\def\eq#1{{Eq.~(\ref{#1})}}
\def\fig#1{{Fig.~\ref{#1}}}

\newcommand{\as}{\alpha_s}

\newcommand{\Lb}{\left(}
\newcommand{\Rb}{\right)}

\parskip=\itemsep               

\newcommand{\D}{\partial}
\newcommand{\h}{\frac{1}{2}}

\newcommand{\f}{\frac}

\newcommand{\la}{\lambda}

\renewcommand{\theequation}{\thesection.\arabic{equation}}

\newcommand{\mL}{\mathcal{L}}
\title{\LARGE \bf Glauber - Gribov approach  for  DIS  on nuclei  in  N=4 SYM }
\author{\large  E. ~Levin\,\, and \,\,J.~Miller   \\
Department of Particle Physics, School of Physics and Astronomy\\
Raymond and Beverly Sackler
 Faculty
of Exact Science\\  Tel Aviv University, Tel Aviv, 69978, Israel\\}

\author{\large
B.Z. Kopeliovich\,\, and\,\,Ivan  \,\, Schmidt\,\\
Departamento de F\'\i sica
y Centro de Estudios
Subat\'omicos,\\ Universidad T\'ecnica
Federico Santa Mar\'\i a, Avda. Espa\~na 1680,\\
Casilla 110-V, Valpara\'\i so, Chile \\
}

\abstract{ In this paper the Glauber-Gribov approach for
deep-inelastic scattering (DIS) with nuclei is developed in N=4 SYM.
It is shown that the amplitude displays the same general properties,
such as geometrical scaling, as is the case in the high density QCD
approach. We found that the quantum effects leading to the graviton
reggeization, give rise to an imaginary part of the nucleon
amplitude, which makes the DIS in N=4 SYM  almost identical to the
one expected in high density QCD.  We concluded that the impact
parameter dependence of  the nucleon amplitude is very essential for
N=4 SYM, and the entire kinematic region can be divided into three
regions which are discussed in the paper. We revisited the dipole
description for DIS and proposed a new renormalized Lagrangian for
the shock wave formalism which reproduces the Glauber-Gribov
approach in a certain kinematic region. However the saturation
momentum turns out to be independent of  energy, as it has been
discussed  by Albacete, Kovchegov and Taliotis. We discuss the
physical meaning of such a saturation momentum $Q_s(A)$ and argue that
one can consider only $Q>Q_s(A)$  within the shock wave approximation.}


\preprint{  TAUP -2889-08\\
USM-TH-241\\
\\hep-ph/???????\\
\today}
\begin{document}

\numberwithin{equation}{section}

\section{Introduction}
The goal of this paper is very modest and pragmatic: to write a
Glauber-type formula for deep inelastic scattering (DIS) with a
nucleus in N=4 SYM. N=4 SYM at weak couplings is similar to our
microscopic theory of QCD, with gauge colour group SU($N_c$). The
high energy amplitude in this theory is given by the exchange of the
BFKL Pomeron, like in QCD \cite{BFKL4}. On the other hand, the
AdS/CFT correspondence \cite{AdS-CFT} allows us to calculate this
amplitude in the strong coupling limit, where it reveals a Regge
behavior (see Ref. \cite{BST1,BST2,BST3} and references therein).
Therefore, in principle, considering the high energy scattering
amplitude in N=4 SYM, we can guess what physics phenomena could be
important in QCD, in the limit of strong coupling.

The simplest and most informative process that allows to study
physics in the region between short distances and long distances, is
DIS in the wide  range of photon virtualities  $Q$. Since the
typical distances are $r \propto 1/Q$, we can approach the long
distance physics at small values of $Q$.  In QCD,  we see three
different regions for DIS:
\begin{enumerate}
\item \quad $Q^2 \gg Q^2_s(x)$ where $Q^2_s(x)$ is the new scale:
saturation momentum (see Refs. \cite{GLR,MUQI,MV} and a  short but
beautiful review in Ref. \cite{MHI}). At such large $Q^2$, we can
use a linear evolution equation, namely the DGLAP equation
\cite{DGLAP}, and the BFKL equation \cite{BFKL}, and all advantages
of the operator Product Expansion \cite{OPE}.
\item \quad $\Lambda^2_{QCD} \,\ll\,Q^2 \,\ll\, Q^2_s(x)$. In this region the density of partons (gluons) is so large that we cannot use here the methods of perturbative QCD. However,  the QCD couplings are still small here, since the typical distances in this kinematic region are $r \propto 1/Q_s(x)$, and $Q_s(x) \gg  \Lambda_{QCD}$ . This fact allows us to suggest a theoretical approach in this region, based on non-linear equations \cite{B,K,JIMWLK}.
\item \quad   $ Q^2 \leq \Lambda^2_{QCD} $.  No rigorous theoretical approach has been developed in this region in QCD. In high energy phenomenology, we describe this region with the soft Pomeron.  However, quite a different phenomenological approach has been tried in this region, namely, that the saturation scale determines the physics inside this domain, and instead of the soft Pomeron, we can use the scattering amplitude in the saturation region (see Refs. \cite{BTAV,KOLE}).  Our intention is to use the input from our N=4 SYM experience, to penetrate this domain.
\end{enumerate}

It turns out that N=4 SYM leads to normal QCD like physics in the
first region, with OPE  and linear equations (see Refs.
\cite{POST}). It has been shown in Ref. \cite{MHI} that the DIS
densities reach saturation at some value of momentum ($Q_s(x)$).
However, the physical picture inside the saturation domain turns out
to be completely different\cite{MHI}, in the sense that there are no
partons in this region and the main contribution stems from
diffractive processes when the target (proton) either remains
intact, or is slightly excited. Such a picture not  only contradicts
the common wisdom, but also contradicts available experimental data.

In this paper we would like to develop a systematic approach to DIS with a nucleus, based on the eikonal  formula.  In QCD the most reliable approach has been developed for this particular case, since a new parameter appears $\as A^{1/3} \approx 1$, which allows to prove the non-linear equation \cite{K}.

\section{Eikonal approximation for scattering with nuclei.}

\subsection{General approach}
It is well known that the  eikonal  approach is based on two main
ideas\cite{Glauber,GRIBGL}. The first one is the fact that the value
of typical  impact parameter for the interaction with a proton is
much smaller than the typical impact parameter  for the nucleon
distribution in a nucleus. Using this idea, we can easily express
the amplitude for interaction with a  nucleus via the interaction
amplitude with a nucleon. Indeed, let us consider a simple example
when the amplitude of interaction with the nucleon is small.
Consider for example deep inelastic scattering (DIS) with a nucleon.
The DIS amplitude for the virtual photon $(\gamma^*)$ interaction
with the nuclear target $( A)$, can be written as follows

\beq \label{GA1} A\Lb \gamma^* A ; s,b\Rb \,\,\,=\,\,\int d^2 b'
\,A\Lb \gamma^* N ; s,b'\Rb\, S\Lb \vec{b} - \vec{b}' \Rb
\,\,\longrightarrow\,\,\int d^2 b' \,A\Lb \gamma^* N ; s,b'\Rb\,S\Lb
b\Rb \eeq
where $S\Lb b \Rb$ is the distribution of nucleons in the nucleus,
normalized as $\int d^2 b  \,S\Lb b \Rb\,\,=\,\,A$, where $A$ is the
number of nucleons in a nucleus. In \eq{GA1} we use the fact that
$|\vec{b} - \vec{b}'|\, \approx\, R_A \,\gg \,R_N \, \approx \,b'$.
$R_A$ is the nucleus radius while $R_N$ is the nucleon size. $\int
d^2 b' \,A\Lb \gamma^* N ; s,b'\Rb$ is equal to the forward
scattering amplitude $A_N(s,t=0)$. In the original Glauber-Gribov
approach it was assumed that $A_N(s,t=0)$ at high energy is mostly
imaginary,  and $ Im A_N(s,t=0) = \sigma^N_{tot}$\footnote{ It
should be noticed that such normalization  of the amplitude is a bit
unusual for high energy physics since the amplitude, calculated from
the Feynman diagrams, has a different normalization, namely, $Im A =
s \,\sigma_{tot}$. We call the first one as non-relativistic while
the amplitude of Feynman diagrams will be called relativistic.}

The second important observation is the fact that at high energies
the longitudinal and transverse degrees of freedom are factorized in
such a way, that in first approximation the interactions with many
nucleons in a nucleus will affect the transverse degrees of freedom
and the impact parameter distribution, but we can neglect the
feedback of these interactions  on the momentum and the trajectory
of the fast projectile. In other words, we can use the eikonal
approximation for high energy scattering.

To illustrate this point, let us consider the interaction of the fast particle with the nucleus at rest, as it is shown in \fig{nuclgl}.

\FIGURE[h]{
\centerline{\epsfig{file=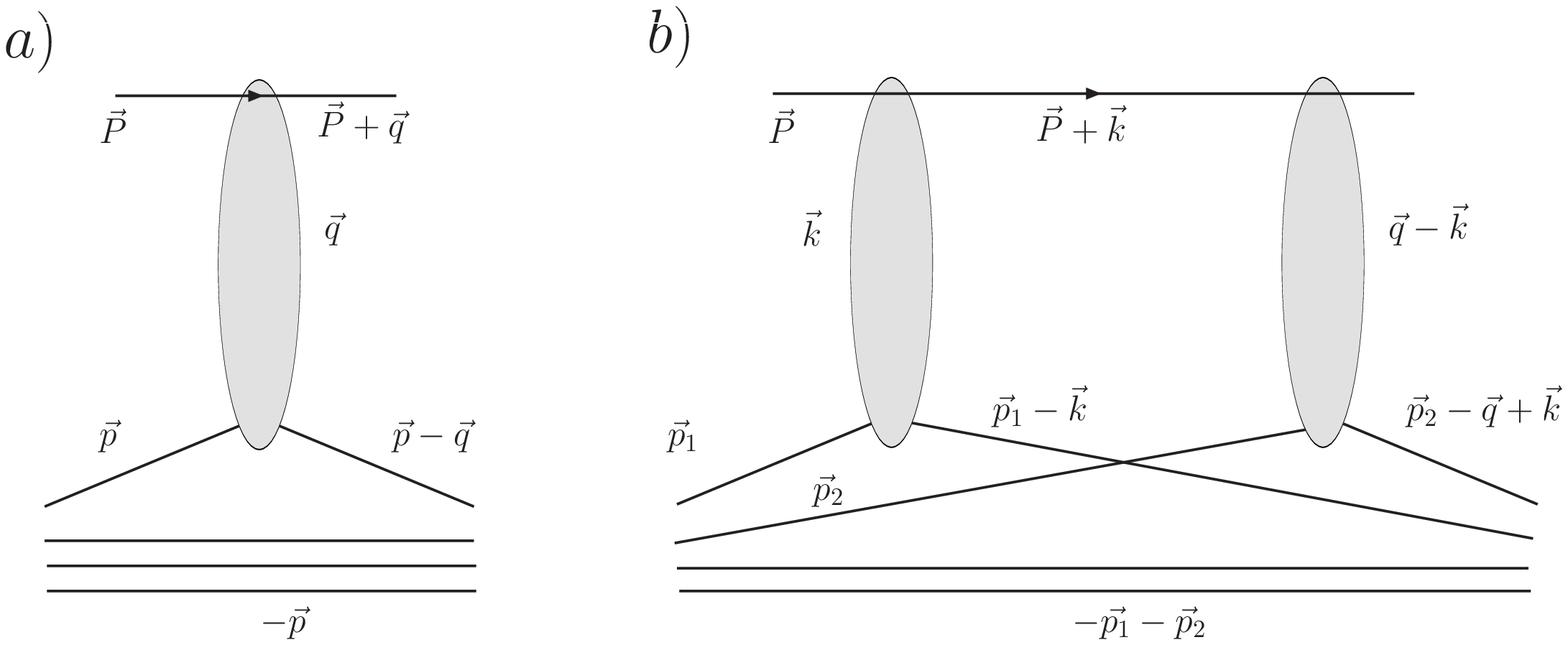,width=140mm,height=40mm}}
\caption{The single (\protect\fig{nuclgl}-a ) and double  (\protect\fig{nuclgl}-b) rescattering with heavy nucleus.
\label{nuclgl} }}

First we demonstrate that the momentum transferred $q$ in \fig{nuclgl}, is transverse at high energy.
For the nuclear target, it is preferable to discuss a process in the rest frame of a nucleus. Describing the nucleus in the non-relativistic approach,  we consider that the kinetic energy of a nucleon is much smaller than  its momentum, namely, $p^2/2 m \,\ll\,|\vec{p}|\,\sim\,1/R_A$ where $R_A \gg R_N$.  Since after rescattering, the nucleon with momentum $\vec{p} - \vec{q}$ is still in the same nucleus,
$q_0 = p_0 - (p - q)_0 = p^2/2m - |\vec{p} - \vec{q}|^2/2m \ll |\vec{q}|$. In our frame, $s = 2 E \,M_A$ where $E$ is the energy of the projectile, and $M_A$ is the mass of the nucleus.  At high energy, the momentum of the projectile is $P = (E,0,0,E)$.  Using the fact that $P^2 = m^2_p$ and $(P + q)^2 = m^2_p$ where $m_p$ is the mass of projectile, we obtain that

\beq \label{GA2}
2 P \cdot q = - q^2\,;\,\,\,\,\,\,\,\,\,\, q_0 \,-\, q_z = -q^2/2E
\eeq
where $z$ is the beam direction. Calculating $q^2$ we  have
\beq \label{GA3}
q^2 = (q_0 + q_z)(q_0 - q_z) - q^2_{\perp} \,=\,-q^2/2E ( 2 q_0  + q^2/2E) \xrightarrow{E \gg  m}
 - q^2_{\perp}
\eeq

The expression for the diagram of \fig{nuclgl}-a has the following form

\bea \label{GA4}
A_A\Lb s, q^2 \Rb\,\,&= &\,\,
\int  \frac{ d^4 p_1}{ ( 2 \pi)^4 i}  \,
\frac{1}{m^2 - p^2 - i\epsilon} \,A_N\Lb s, q^2_{\perp}; p^2_1,(p - q)^2\Rb; \,\frac{1}{ m^2 - (p - q)^2 - i\epsilon}\,\nonumber\\
& &\int \,\prod^{A-1}_{i=1}\frac{ d^4 p_i}{ ( 2 \pi)^4 i}\, \Gamma\Lb p_1;\{p_i\}\Rb\frac{1}{m^2 - p^2_i - i \epsilon}
 \,\Gamma\Lb p _1 - q;\{p_i\}\Rb
\eea
where $\Gamma\Lb p_1;\{p_i\}\Rb $ is the vertex for the transition
of the nucleus into $A$ free nucleons. Introducing a new variable
for the energies of the nucleons, namely, $p_{0,i} \,\equiv\,M_A/A -
\tilde{p}_{0,i}$ and noticing that since $\tilde{p}_{0,i}$ has the
interpretation of being the kinetic energy, we anticipate very small
values of $\tilde{p}_{0,i}\,\, \ll\,\, |\vec{p}_i|$, and therefore
we can neglect $\tilde{p}^2_{0,i}$ . Using this approach, each
propagator has the form

\bea \label{GA5}
&&m^2 - p^2_i - i\epsilon \,\,=\,\,(-\,\frac{M^2_A}{A^2} +m^2) \,\,+\,\,2 \tilde{p}_{0,i}\,\frac{M_A}{A}\,\,  +\,\, |\vec{p}_i|^2\,\, -\,\, i \epsilon \,\,=\,\,m\,{\Large\varepsilon} +2 \tilde{p}_{0,i}\,m\,   +\,\, |\vec{p}_i|^2\,\,  - i \epsilon
\,\,\,\,\mbox{for $ i\,< \,A$}\,\,\, \nonumber\\
&&\mbox{but}\,\,m^2 - p^2 - i\epsilon \,=\,  (-\,\frac{M^2_A}{A^2}
+m^2) -2 \sum^{A-1}_{i=1}\,\tilde{p}_{0,i}\,\frac{M_A}{A}
\,+\,|\vec{p}|^2 - i \epsilon  \,=\,m\,{\Large\varepsilon}\,\,-2\,\,
\sum^{A-1}_{i=1}\,\tilde{p}_{0,i}\,m\,+\,|\vec{p}|^2\,  -\, i
\epsilon \eea
where ${\Large \varepsilon} = (M_A - A m)/A$ is the
bounding energy per one nucleon in a nucleus, which is much smaller
than the mass of the lightest hadron. One can see that all
propagators for $i\,<\,A$, have poles in $
 \tilde{p}_{0,i}$ in the upper semi-plane, while the $A$-th propagator has a pole in the lower semi-plane. Closing the contour of integration over $ \tilde{p}_{0,i}$, on the poles in the lower semi-plane, we obtain the following anticipated result, namely

\bea \label{GA6}
A_A\Lb s, q^2 \Rb\,\,&= &\,\,
\int \,\prod^A_{i=1} \frac{ d^3 p_i}{ ( 2 \pi)^3}  \, \Gamma\Lb p_1;\{p_i\}\Rb
\frac{1}{ A\,{\Large  \varepsilon} -  \,\sum^A_{i=1} \,\frac{|\vec{p}_i|^2}{2 m} - i\epsilon} \,A_N\Lb s,  q^2_{\perp};p^2_1,(p - q)^2\Rb\,\nonumber \\
 & \times & \,\frac{1}{ A\,{\Large \varepsilon}  -\frac{ (\vec{p}_1 - \vec{q})^2}{2 m} - \,\sum^A_{i=2} \,\frac{|\vec{p}_i|^2}{2 m}\, i\epsilon}
 \,\Gamma\Lb p _1 - q;\{p_i\}\Rb
\eea

The above calculation did not take into account the possible singularities in the nucleon amplitude, since their positions are determined by the mass of hadrons $ \tilde{p}_{0,i} \approx\, m_\pi$. Closing the contour on these singularities, we obtain a smaller contribution of the order of $1/m_\pi R_A$.\\

Introducing the wave function of the nucleus as follows

\beq \label{GA7}
\Psi\Lb \{r_i\}\Rb \,\,=\,\,\int \,\prod^A_{i=1} \frac{ d^3 p_i}{ ( 2 \pi)^3} e^{i \vec{p}_i \cdot \vec{r}_i }\, \Gamma\Lb p_1;\{p_i\}\Rb
\frac{1}{ A\,{\Large \varepsilon }-  \,\sum^A_{i=1} \,\frac{|\vec{p}_i|^2}{2 m} - i\epsilon}
\eeq
we can rewrite \eq{GA5} in the form

\bea \label{GA8}
A_A\Lb s, q^2 ; \fig{nuclgl}-a\Rb\,\,\,&=& \,\,\,A_N \Lb s,q^2_{\perp}\Rb\,\,\int\,\,\prod^A_{i=1}\,d^3 \,r_i\,e^{ i \vec{q}_{\perp} \cdot \vec{r}_{1,\perp}} \,|  \Psi\Lb \{r_i\}\Rb |^2\,\, \\
 &\rightarrow & \,\,
A_N \Lb s,q^2_{\perp}=0\Rb  \int\,\,\prod^A_{i=1}\,d^3 \,r_i\,e^{ i \vec{q}_{\perp} \cdot \vec{r}_{1,\perp}} \,|  \Psi\Lb \{r_i\}\Rb |^2 \, \equiv\,A_N \Lb s,q^2_{\perp}=0\Rb \,S\Lb q^2_\perp\Rb
\nonumber
\eea
which  is  \eq{GA1} in momentum representation. In deriving \eq{GA8}, we used the fact that in
$S\Lb q^2_\perp\Rb$, the typical $q_{\perp} \propto 1/R_A$,  which is much smaller than
the characteristic $q_{\perp}$ in the nucleon amplitude, and which can be considered to be a constant as far as the $q_{\perp}$ dependence is concerned. Now we want to show that the diagram of \fig{nuclgl}-b leads to the following contribution

\beq \label{GA9}
A_A\Lb s, b; \fig{nuclgl}-b\Rb\,\,\,=\,\,\,i\,\frac{1}{2} \,\Lb\int\,d^2 b' A_N\Lb s,b'\Rb\Rb^2\,S^2\Lb b \Rb
\eeq
It turns out that \eq{GA9} can be obtained with the additional assumption that the wave function
can be factorized as
\beq \label{GA10}
\Psi\Lb \{r_i\}\Rb \,\,=\,\,\prod^A_{i=1}\,\,\Psi\Lb r_i\Rb \,\,\mbox{,  which gives  }\,\,
S\Lb b\Rb = \int\,d z\, |\Psi\Lb b,z\Rb|^2 \,\,\,\mbox{,  with} \,\,\vec{r} \,= ( \vec{b}_{\perp}, z ).
\eeq
This means that there are no correlations between different nucleons in a nucleus. In other words, we describe the nucleus as the nucleons that are moving in the external potential in the spirit of the Hartree-Fock approach.

The amplitude for  the diagram of \fig{nuclgl}-b has the form

\bea \label{GA11}
A_A\Lb s,q^2;\fig{nuclgl}-b \Rb \,\,&=&\,\,\int \frac{ d^4 k}{(2 \pi )^4 i}\,\frac{1}{ m^2_p - (P + k)^2}
\int \frac{ d^4 p_1}{ ( 2 \pi)^4 i} \,\frac{ d^4 p_2}{ ( 2 \pi)^4 i}\prod^A_{i=3}\,\frac{ d^4 p_i}{ ( 2 \pi)^4 i}  \, \,\Gamma\Lb p_1,p_2, \left\{ p_i\right\}\Rb\nonumber\\
 &\times&\, \,\frac{1}{m^2 - p^2_2- i\epsilon}
\frac{1}{m^2 - p^2_i1- i\epsilon} \,A_N\Lb s, k^2_{\perp}; p^2_1,(p_1 - k)^2\Rb; \nonumber\\
&\times &\frac{1}{m^2 - (p_1 - k)^2 - i\epsilon}\,\,A_N\Lb s, ( q -k)^2_{\perp}; p^2_2,(p_2 -q + k)^2\Rb
\nonumber\\
 & \times &\,\,
\frac{1}{m^2 - (p_2 - q  +   k)^2 - i\epsilon} \,\frac{1}{m^2 - p^2_i - i\epsilon}
\,\Gamma\Lb p_1-k,p_2-q+k,\left\{ p_i\right\}\Rb
\eea

We integrate first over the momentum $k$. Rewriting $d^4 k$ as $ d k_0 d  (k_0 - k_z) d^2 k$,  and closing the contour of integration over the variable $k_0 - k_z$,  on the pole $(P + k)^2 = m^2_p$, leads to a factor of $2 \pi i /P_0$.
For the integration over $k_0$, we can also close the contour on one of the poles: $(p_1 -k)^2= m^2$
 or $(p_2 - q +k)^2=m^2$, which can be rewritten as $m\, {\Large \varepsilon} + 2 m\, (\tilde{ p}_{0,1} - k_0)  - (\vec{p}_1 - \vec{k})^2 - i \epsilon=0$ and $ m\, {\Large \varepsilon} + 2 m\, (\tilde{ p}_{0,2} - q_0 + k_0)  - (\vec{p}_2 - \vec{q} + \vec{k})^2$.  This integration brings an additional factor of $2 \pi i/2 m$. Therefore, the integration over $k$ leads to  the following contribution, namely $ i\,d^2 k/((2 \pi)^2\,s)$.
 Evaluating all the integrations over $\tilde{p}_{0,i}$ in the same way as we did when calculating the diagram of \fig{nuclgl}-a, we reduce \eq{GA11} to the following expression

\bea \label{GA12}
A_A\Lb s, q^2_\perp; \fig{nuclgl}-b\Rb &=& \frac{i}{s}\,\,\int \,\frac{d^2 k}{(2 \pi)^2}\,
\int \,\prod^A_{i=1} \frac{ d^3 p_i}{ ( 2 \pi)^3}  \, \Gamma\Lb p_1;\{p_i\}\Rb
\frac{1}{ A\,{\Large  \varepsilon} -  \,\sum^A_{i=1} \,\frac{|\vec{p}_i|^2}{2 m} - i\epsilon} \,A_N\Lb s,  k^2_{\perp};p^2_1\Rb\, \\
 & \times & A_N\Lb s, ( q - k)^2_{\perp};p^2_1\Rb\frac{1}{ A\,{\Large \varepsilon}
 -\frac{ (\vec{p}_1 - \vec{k})^2}{2 m} - \frac{ (\vec{p}_2 -  \vec{q} + \vec{k})^2}{2 m} \,-\,\sum^A_{i=3} \,\frac{|\vec{p}_i|^2}{2 m}\, i\epsilon}
 \,\Gamma\Lb p _1 - q;\{p_i\}\Rb \nonumber
\eea

\eq{GA12} can be easily rewritten in coordinate representation, by introducing the wave function of \eq{GA7}, namely

\beq \label{GA13}
A_A\Lb s, b; \fig{nuclgl}-b\Rb\,=\,\frac{i}{s} A^2_N\Lb s, q^2 =0\Rb\,\int d z_1 \int^{z_1} d z_2
\,\,|\Psi\Lb, b,z_1; b, z_2;\{r_i\}\Rb|^2
\eeq

Using the non-relativistic normalization for the scattering amplitude ( $A_{nr} = A/s$)\footnote{Starting from this equation we will use the notation $A_N$ and $A_A$  for the non-relativistically normalized amplitudes, hoping that it will not lead to any misunderstanding.} and \eq{GA10}, we can see that we
obtain \eq{GA9}. It should be noted that the factor $1/2$ stems from the $z_2$ integration, which is not restricted in \eq{GA9}, in contrast with \eq{GA13}.
All calculations above have been done to illustrate two points, namely that we do not need to assume that the nucleon amplitude should be pure imaginary, but we need to assume a very simple model for the nuclei.

Calculating the amplitude for the interaction with any number of nucleons in a nucleus, we obtain the simple formula for the nucleus scattering amplitude (see a more detailed derivation in Ref. \cite{GRIBGL}), namely,

\beq \label{GA14} A_A\Lb s,b\Rb \,\,=\,\,i \Lb\,1\,\,\,-\,\,\,\exp
\Lb i\, \int d b'  A_N(s, b')\,S\Lb b\Rb\Rb\Rb \eeq
In deriving \eq{GA1} we considered the propagators of the projectile
and the target (nucleons in a nucleus) in flat space but not in
$AdS_{5}$. In the next section we will comment on this but the main
argument is very simple: the  trajectory of a fast moving particle
can be replaced by the straight line in curved space as well as in
flat one. The second assumption was that we considered in
\fig{nuclgl}-b the projectile as the intermediate state.

\FIGURE[h]{\begin{minipage}{80mm}
\centerline{\epsfig{file=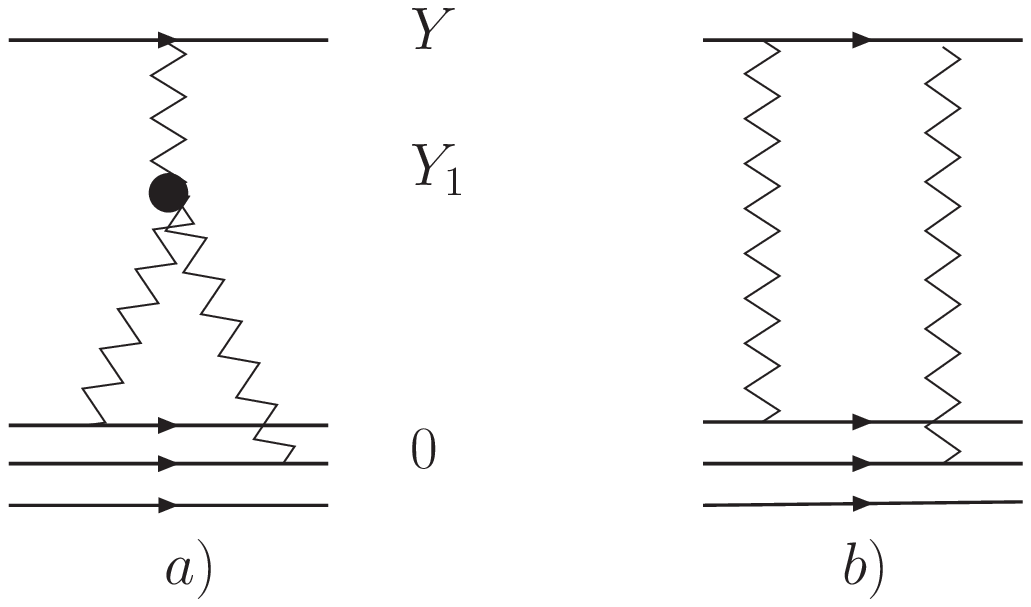,width=70mm,height=30mm}}
\caption{ The first fan diagram (\fig{3pom}-a) for the interaction of Pomerons  (reggeized gravitons) and the eikonal diagram (\fig{3pom}-b).}
\end{minipage}
\label{3pom}}

Using the AdS/CFT correspondence we can estimate the accuracy of
this (eikonal)  approach in the N=4 SYM case.  Indeed, at first
sight we can expect  from  the AdS/CFT correspondence, that the main
contribution will stem from the fan diagrams, the first of which is
shown in \fig{3pom}-a, as it happens in this theory in the   region
of small coupling constant. In fact, from the region of small
coupling we expect that (i) this diagram has the contribution of the
order of $(\as^5/\Delta)\,s^{2 \Delta}$, where $\Delta \propto \as$
is the intercept of the BFKL Pomeron; (ii) the typical value of $Y -
Y_1 \,\approx\,1/\Delta\,\gg\,1$ and (iii) the value of this
contribution is closely related to the process of diffractive
dissociation of the projectile. Since $Y - Y_1\,\gg\,1$ it is
reasonable to consider the exchange of the BFKL Pomeron. The eikonal
diagram of \fig{3pom}-b has the same order of magnitude but  it
turns out (see Ref.\cite{BAP}) that this diagram is included in the
diagram of \fig{3pom}-a in the region of integration $Y - Y_1
\,\approx\,1$ where we cannot use Pomeron exchange. Therefore, in
the weak coupling limit the full set of diagrams at high energy can
be reduced to the ''fan'' diagrams. It is worth  mentioning that in
the weak coupling limit the eikonal diagram of \fig{3pom}-b has the
same intermediate state as the initial one  (the colourless dipole)
since it turns out that colourless dipoles are diagonalized by the
interaction matrix (see Ref. \cite{MUCD}).

In the strong coupling limit of N=4 SYM, due to the AdS/CFT
correspondence, the strong interaction of Pomerons is replaced by
the weak interaction of the reggeized gravitons, with intercepts
$\Delta = 1 - 2/\sqrt{\lambda}$, and therefore in the triple Pomeron
diagram the typical value of $Y - Y_1 \,\approx 1/(\Delta = 1 -
2/\sqrt{\lambda}) \approx 1$. It means that diffraction production,
which can contribute and was neglected in the eikonal
(Glauber-Gribov)  approach , is the process in which low masses are
produced.  For $Y - Y_1\, \approx\,1$ there are no reasons to
replace the amplitude by the  reggeized graviton exchange.  Using
the AdS/CFT correspondence we expect that in the diagram of
\fig{3pom}-b the  same as the initial state is produced. On the
other hand, the process of diffraction production of low mass can be
easily taken into account in the eikonal approach, and  does not
change neither the energy nor the impact parameter dependence that
has been discussed here. The cross section of the diffraction
dissociation is proportional to the imaginary part of the  reggeized
graviton exchange which is small of the order of $2/\sqrt{\lambda}$.
Therefore, at least within this accuracy ( $2/\sqrt{\lambda}$), the
exchange of two gravitons between the projectile and the target
(eikonal diagram of \fig{3pom}-b) prevails.

\subsection{Nucleon amplitude in N=4 SYM}

The main contribution to the scattering amplitude at high energy  in
N=4 SYM, stems from the exchange of the graviton\footnote{Actually,
the graviton in this theory is reggeized \cite{BST1}, but it is easy
to take this effect into account (see Refs. \cite{BST1,BST3,MHI}).}
. The  formula for this exchange has been written in
Ref.\cite{BST2,COCO}, (see also Ref. \cite{MHI} for its interpretation).
In flat space this amplitude has the following form

\beq \label{N41}
A_g(s,q)\,\,\propto\,\,T_{\mu\nu}\Lb p_1,p_2\Rb G_{\mu \nu \mu^{\,\prime} \nu^{\,\prime}}\Lb q\Rb\,T_{\mu^{\,\prime}\nu^{\,\prime}}\Lb p_1,p_2\Rb\,\, \xrightarrow{s\gg q^2} s^2 \,\frac{1}{q^2_\perp}
\eeq

where $T_{\mu,\nu}$ is the energy-momentum tensor, and $G$ is the
propagator of the massless graviton. The last expression in
\eq{N41}, stems from the fact that for high energies, $T_{\mu,\nu} =
p_{1,\mu} p_{1,\nu}$, and  $q^2 = - q^2_\perp$ (see the previous
section). However, we are interested in N=4 SYM in a space with
curvature, namely $ AdS_5$.
 $AdS_{d + 1}$ corresponds to an hyperboloid in $ d + 2$ flat space, namely

\beq \label{N42}
 - Y^2_{-1} \,+\,Y^2_0 \,+\,\sum^d_{I=1} \,X^2_i= - L^2
\eeq
with curvature $R=- d(d -1) L^2$. Introducing new coordinates

\beq \label{N43}
x_i\,\,=\,\,\frac{L\,X_i}{Y_0 + Y_{-1}}\,;\qquad z\,\,=\,\,\frac{L^2}{Y_0 + Y_{-1}}\,;
\eeq
we reduce the introduced metric to the following form

\beq \label{N44}
d s^2\,\,=\,\,\frac{L^2}{z^2}\,\Lb \,d z^2\,\,+\,\,\sum^d_{i=1}  d x^2_i \Rb\,=\,\frac{L^2}{z^2}\,\Lb \,d z^2\,+\,d \vec{x}^2 \Rb
\eeq

In the flat $d + 2$ dimensional space, the scalar propagator  is the following ( with $Y_+ = Y_0 + Y_{-1}$ and  $Y_- = Y_0 - Y_{-1}$)

\bea \label{N45}
G \Lb  X_i,Y_+,Y_-; X'_i,Y'_+,Y'_- \Rb\,\,&=&\,\,\int \prod^d_{i=1}\frac{ d k_i}{2 \pi}\,\frac{d p_{+} d p_{-}}{(2 \pi)^2}\,\frac{1}{\sum^d_{i=1}  k^2_i\,+\,p_+\,p_-}\,\,e^{ - i \vec{k}\cdot\vec{X}  - i \h p_+ Y_- - i \h  p_- Y_+} \nonumber \\
 &=& \int^{\infty}_0\,d t \int \prod^d_{i=1}\frac{ d k_i}{2 \pi}\,\frac{d p_{+} d p_{-}}{(2 \pi)^2}\,\exp\Lb - t\, k^2 - t\,p_+p_- -  i \vec{k}\cdot\vec{X}  - i \h p_+  Y_-  - i \h  p_-  Y_+\Rb \nonumber\\
&=& (2 \pi)^{-d/2 -1}\,\int^\infty_0 \,d t\, t^{-d/2 - 1} \,e^{-u/t}\,\xrightarrow{t \to 1/\xi} \,\,(2 \pi)^{-d/2 -1}\,\int^\infty_0 \,d \xi  (\xi)^{d/2 -1}\,e^{ - \xi\,u} \,\,\nonumber \\
&=& \,\,(2 \pi)^{-d/2 -1}\,\Gamma\Lb  d/2\Rb \,u^{ - \h d}
\eea

In \eq{N45}, $u$ is a new variable which is equal to

\beq \label{u}
u\,\,=\,\,\frac{ (z - z')^2 + (\vec{x} - \vec{x}')^2}{ 2 \,z\,z'}.
\eeq

In \eq{N45}, we re-write the integration measure of the momenta in  $d + 2$  dimensional space, namely $\prod^{d + 2}_{i=1}  d p_i$, as $ \prod^d_{i=1} d k_i \,d p_+ d p_- $,  where $p_+$ and $ p_-$ are the conjugated momenta to $Y_-$ and $Y_+$, respectively.

However, the propagator of \eq{N45} does not satisfy  the correct
boundary condition, for example, $G \Lb  X_i,Y_+,Y_-; X'_i,Y'_+,Y'_-
\Rb$ should approach $\delta\Lb\vec{X} - \vec{X}'\Rb $ as $z \to
z'$, which is not the case for this equation. One of the reasons why
this happens, is that we have to guarantee that $Y_+ >0$\footnote{We
thank Chung-I Tan for the fruitful discussion of all aspects of high
energy scattering in N=4 SYM,  in particular, the $Y_+ >0$
condition.}.  The easiest way to impose such a condition, is to
change \eq{N45} to

\bea \label{N450}
G \Lb  X_i,Y_+,Y_-; X'_i,Y'_+,Y'_- \Rb\,\,&=&\,\,\int \prod^d_{i=1}\frac{ d k_i}{2 \pi}\,
\frac{d p_{+} d p_{-}}{(2 \pi)^2}\,\frac{1}{\sum^d_{i=1}  k^2_i\,+\,p_+\,p_-}\,\,\frac{1}{p_-}\,e^{ - i \vec{k}\cdot\vec{X}  - i \h p_+ Y_- - i \h  p_- Y_+} \nonumber \\
&=& \,\,\int \prod^d_{i=1}\,\frac{ d k_i}{2 \pi}\,\frac{d p_{+} }{(2 \pi)^2}\,\frac{1}{k^2}\,\left\{
e^{i\,\frac{k^2}{p_+}\,Y_+} \,\,-\,\,1 \right\}\,\,e^{ - i \vec{k}\cdot\vec{X}  - i \h p_{+} Y_-}
\eea
One can see from \eq{N450} that $\Lb \sum^{i=3}_{i=0} \partial^2/\partial^2 X_i
 + \partial^2/\partial^2 Y_0  - \partial^2/\partial^2 Y_{-1}\Rb G \Lb  X_i,Y_+,Y_-; X'_i,Y'_+,Y'_- \Rb$ is defined only for
$Y_+ >0$. Therefore, the solution of the equation for the Green's function also will be determined only for $Y_+ >0$.

Notice that the mass of the graviton is equal to zero even in the  $AdS_{d + 1}$ space with curvature. Having this in mind, the easiest way to find the correct propagator, is to write the wave equation directly in the $AdS_{d+1}$ space, assuming that the mass of the graviton is equal to zero, and
that $G \Lb  X_i,Y_+,Y_-; X'_i,Y'_+,Y'_- \Rb $ is a function of the variable $u$ of \eq{u}. The action  for such a particle has the following form

\beq \label{S}
S[\phi] = \h \int d^d x \,d z \,\sqrt{g}\, g^{\mu, \nu} \,\partial_{\mu} \phi\,\partial_{\nu} \phi
\eeq
where  the metric is given by \eq{N44} . Using \eq{S} and \eq{N44}, it is easy to obtain the wave equation for $G \Lb  X_i,Y_+,Y_-; X'_i,Y'_+,Y'_- \Rb\,\,\,=\,\,\,G\Lb u\Rb$. It has the form\cite{GRF4,GRF4M}

\bea
&&\frac{1}{\sqrt{g}}\,\partial_{\mu} \sqrt{g} \,g^{\mu.\nu}\partial_{\nu}\,G(u)\,\,=\,\,0 \,;\label{N461}\\
&&z^2 \nabla^2_{x} G(u)  \,\,+\,\,z^{d + 1}\frac{\partial}{\partial z}\,\left[z^{- d + 1}\, \frac{\partial G(u)}{\partial z}\,\right]\,\,=\,\,0\,; \label{N462}\\
&& u\,(u + 2)\,G_{u,u}\Lb u \Rb\,\,+\,\,(d + 1 )\,G_u \Lb u \Rb\,\,\,=\,\,0\,; \label{N463}
\eea

The solution to \eq{N463}, which satisfies all the necessary boundary conditions:
$G\Lb u\Rb \xrightarrow{u \to \infty} \,0$ and $G\Lb u\Rb \,\xrightarrow{ z \to z'} \,\delta\Lb \vec{x} - \vec{x}' \Rb$ has the form\cite{GRF4,GRF4M,COCO}

\beq \label{N47}
G\Lb u \Rb \,\,\,=\,\,\,\frac{ d -1}{2^{d +1}}\,\Lb \frac{1}{4 \pi}\Rb^{\h d}\,\,\Lb - \frac{2}{u} \Rb^d\,\, {}_2F_1\Lb d, \h(d + 1),d + 1,- \frac{2}{u} \Rb
\eeq

As has been discussed (see \eq{N41}), we need an expression for the
propagator of the graviton,  which at high energy depends only on
the transverse coordinates for the scattering. Therefore, we need
$G\Lb u \Rb$ for $AdS_{2 + 1}$, which is equal to

\beq \label{G}
G_{3}\Lb u \Rb \,\,=\,\,\frac{1}{4 \pi}\,\frac{1}{\left\{ 1 + u + \sqrt{u (u + 2)}\right\}^2\,\sqrt{u (u + 2)}}
\eeq
with

\beq \label{UTR}
u\,\,=\,\,\frac{ (z - z')^2 + b^2}{ 2 \,z\,z'}
\eeq
where $b$ is the impact parameter for the scattering amplitude.

For the  eikonal  formula, we need to evaluate the integral over
$b$, which can be easily done noticing that

\beq \label{DB}
d \left\{1 + u + \sqrt{u (u + 2)}
\right\}/d b^2\,\, =\,\,\frac{1}{2 \,z\,z'} \,\frac{ \left\{1 + u + \sqrt{u (u + 2)}\right\}}{\sqrt{u (u + 2)}}
\eeq

 The result is

\bea \label{GINT}
&&G(z,z')\,\,= \\
&&\int d^2 b \,G_{3}\Lb u \Rb \,\,=\,\,\frac{z\,z'}{4}\, \frac{1}{ \left\{  1 + u(b=0) +
 \sqrt{u(b=0) (u(b=0) + 2)}\right\}^2}\,\,=\,\,z\,z'\,\frac{z^2\,z'^2}{(z^2 + z'^2 \,+\,| z^2 - z'^2|)^2}\nonumber
\eea

This equation provides us with the factor which enters into
\eq{N41}, instead of $1/q^2_\perp$. It turns out that in curved
space we need to change \cite{BST2}

\beq \label{SRTIL}
s \,\,\,\rightarrow\,\,\,\tilde{s}\,\,=\,\,\frac{s}{\sqrt{g_{+-} (z)\,g_{- +}(z')}} \,\,=\,\,\frac{z z' s}{R^2}.
\eeq

For calculating the nucleon amplitude, we need to multiply \eq{GINT} by the coupling constant, and integrate over  the nucleon wave function \cite{BST2,COCO}. Therefore, the nucleon amplitude is equal to

\bea \label{N48}
  \int d^2 b\, A_N(s,b)\,\,=\,\,i\,g^2_0\,s \int d z' \,z z'\,G(z,z')\,|\Phi(z')|^2\,\,&=&\,\,
i\, g^2_0\,s\,\,\int\,d z' \,|\Phi\Lb z'\Rb|^2\,z'^2\,\frac{z^2\,z'^2}{(z^2 + z'^2 \,+\,| z^2 - z'^2|)^2}\,\,\nonumber\\
 &\xrightarrow{z \ll z'}&\,\,\frac{i\,g^2_0\,s}{4}\,z^4 \,\int\,d z' |\Phi\Lb z'\Rb|^2\,\,=\,\,\frac{i\,g^2_0\,N_c\,s}{4}\,\,\,z^4
\eea
Here $g^2_0$ is the dimensionless constant, which is equal to
$\kappa^2_5/2 L^3$, where $\kappa_5$ is the five dimensional
gravity. $g^2_0\,\, \propto\,\, 1/N^2_c$ where $N_c$ is the number
of colours.  We do not know anything about the nucleon wave
function, except that the integral over $z'$ converges, and it is
proportional to $N_c$. Therefore, the amplitude is proportional to
$A_N \propto s/N_c$ and it is small for $s z^2 < N_c$ .  It grows
and becomes of the order of 1 due to the reggeization of the
graviton. The graviton propagator in \eq{N48} should be replaced by
the propagator of the Pomeron, in the way as has been suggested in
Refs. \cite{BST1,MHI,BST3}.  This modification for our case is
described in section 5.  In \eq{N48}, we consider $\int \,d z'
|\Phi{z')}^2 = N_c $ .

As has been discussed, we use the propagator for a fast moving particle in the form
\beq \label{N49}
G\Lb k_+,k_-; \vec{b}_1 - \vec{b}_2; z_1 - z_2\Rb\,\,=\,\,\frac{1}{k_+(k_ +\,+\, i \epsilon)}\,\delta^{(2)}\Lb \vec{b}_1 - \vec{b}_2\Rb\,\delta \Lb z_1 - z_2\Rb
\eeq
\eq{N49} follows directly from \eq{N450} . Indeed for large $k_+$  the pole in  the integrant of \eq{N450} is located at $k_-\,=\,(k^2_{\perp}  - p_+p_- - i \epsilon)/k_+\,\to\,0  - i \epsilon$. Therefore,  $\sum^{d }_{i=1} k^2_i + p_+p_-$ can be replaced by
$k_+(k_- + i \epsilon)$.  Substituting this expression in \eq{N450} one can see that $G\Lb k_+,k_-; \vec{b}_1 - \vec{b}_2; z_1 - z_2\Rb$ has the form of \eq{N49} with an  additional factor $\Theta ( z_1 + z_2)$ which is equal to 1.

\eq{N49} for $ G\Lb k_+,k_-; \vec{b}_1 - \vec{b}_2; z_1 - z_2\Rb$ can be derived directly from \eq{N461} and \eq{N462}. Indeed, going to Fourier transform for coordinates $x_i\,(i = 1 , \dots,d)$ and to Laplace transform for coordinate $z$ we can rewrite \eq{N462} in the form
\beq \label{N50}
k^2\,\tilde{G}'_p(\{k_i\};p)\,\,-\,\,(d -1)\,p\,\tilde{G}\{k_i\};p)\,\,- \Lb p^2\,\tilde{G}(\{k_i\};p)\Rb'_p\,\,=\,\,0
\eeq
The solution to this equation has the form
\beq \label{N51}
\tilde{G}(\{k_i\};p)\,\,=\,\,\frac{1}{k^2 - p^2}\,\Lb \frac{k^2}{k^2 - p^2}\Rb^{\frac{d - 1}{2}}\,\,=\,\,\frac{1}{k_+\,k_-\,-k^2_{\perp} - p^2}\,\Lb \frac{k_+ k_-\,-\,k^2_{\perp}}{k_+ k_-\,-\,k^2_{\perp}\, -\, p^2}\Rb^{\frac{d - 1}{2}}
\eeq
For large $k_+$  \eq{N51} leads to
\beq \label{N52}
\tilde{G}(\{k_i\};p)\,\,\xrightarrow{k_+\,\ll\,\{k_{\perp} \,\mbox{and}\,\,p\}}\,\,\frac{1}{k _+\,(k_-\,\,-\,\,i\,\epsilon)}
\eeq
\eq{N52} gives \eq{N49} which we use in our calculations.

\subsection{Eikonal formula in N=4 SYM}
\eq{GA14} can be easily rewritten for the case of N=4 SYM in the following way using \eq{N48}

\beq \label{GLF1}
A_A(s,b)\,\,=\,\,i\,\int \,d\,z \,|\Phi_p(z)|^2\,\left\{\,1\,\,-\,\,e^{ i\,{\Large s}\, \frac{\,g^2_0\,N_c}{4}\,\,\,z^4\,
S\Lb b\Rb} \right\}
\eeq

where $\Phi_p$ is the wave function of the projectile. This formula is almost the same as the eikonal formula for  the hadron-nucleus interaction, except that the nucleon amplitude is purely real in our case.

The  scattering amplitude at fixed $z$

\beq \label{GLF2}
A_A\Lb s,b;z\Rb\,\,\,=\,\,i\, \left\{\,1\,\,-\,\,e^{ i\,{\Large s}\, \frac{\,g^2_0\,N_c}{4}\,\,\,z^4
S\Lb b\Rb} \right\}
\eeq
can be rewritten in the following way:

\beq \label{GLF3}
A_A\Lb s,b;z\Rb\,\,\,=\,\,\,\sin\left[ s\, \frac{\,g^2_0\,N_c}{4}\,\,\,z^4\,S\Lb b\Rb \right]\,\,+\,\,i\,\,
2 \sin^2\left[s\, \frac{\,g^2_0\,N_c}{8}\,\,z^4\,S\Lb b\Rb \right]
\eeq

One can see that the real and imaginary part of the amplitude are of the same order in contrast with the black disc behavior, for which only the imaginary part survives at high energy.  One can see that
the amplitude of \eq{GLF3} satisfies the following unitarity constraint

\beq \label{GLF4}
2 Im A_A\Lb s,b;z\Rb\,\,=\,\,|A_A\Lb s,b;z\Rb|^2
\eeq

Comparing \eq{GLF4} with the general unitarity constraint, namely,
$$ 2 Im A\Lb s,b;z\Rb\,\,=\,\,|A\Lb s,b;z\Rb|^2\,\,+\,\,G_{inel}\Lb s,b;z\Rb$$
one can see that \eq{GLF3} leads to only elastic scattering at high energy, in direct contradiction with our intuition based on the parton approach.

For the general formula of \eq{GLF1},  \eq{GLF4} means that

\bea \label{GLF5}
\sigma_{tot} \,&=&\,2 \int d^2b\,\int \,d\,z \,|\Phi_p(z)|^2\,Re \left\{\,1\,\,-\,\,e^{ i\,{\Large s}\, \frac{\,g^2_0\,N_c}{4}\,\,z^4
S\Lb b\Rb} \right\} \,\,=\nonumber \\
\sigma_{diff} \,+\,\sigma_{el} &=& \int d^2b\,\int \,d\,z \,|\Phi_p(z)|^2\,\, \left|\,1\,\,-\,\,e^{ i\,{\Large s}\, \frac{\,g^2_0\,N_c}{4}\,\,\,z^4\,\,
S\Lb b\Rb} \right|^2
\eea

In other words, only the processes of diffractive dissociation contribute at high energy.

\section{DIS with nuclei: general formulae}
For calculating DIS, we need to specify the wave function of the projectile in \eq{GLF3}.
In N=4 SYM, the natural probe for DIS is ${\cal R}$-current (${\cal R}$-boson) \cite{POST}, and the wave function for this probe satisfies  \eq{N462}. However, in DIS we fix the virtuality of the probe (see \fig{disn}). It means that in terms of  \eq{N45}, $\sum^d_{i=1}\,k^2_i = - Q^2$. Therefore, the wave function is described by     \eq{N462} with $ d =0$, but with $\nabla^2_x \Psi = - Q^2 \Psi$, and the equation  can be rewritten in the form \cite{POST}

\beq \label{DIS1}
- z^2 \,Q^2 \Psi_{\cal R}\Lb Q^2,z\Rb  + z \,\frac{ d \Psi_{\cal R}\Lb Q^2,z\Rb}{ d z} \,+\,
z^2\,\frac{ d^2 \Psi_{\cal R}\Lb Q^2,z\Rb}{ (d z)^2}\,\,=\,\,0
\eeq

\FIGURE[t]{
\centerline{\epsfig{file=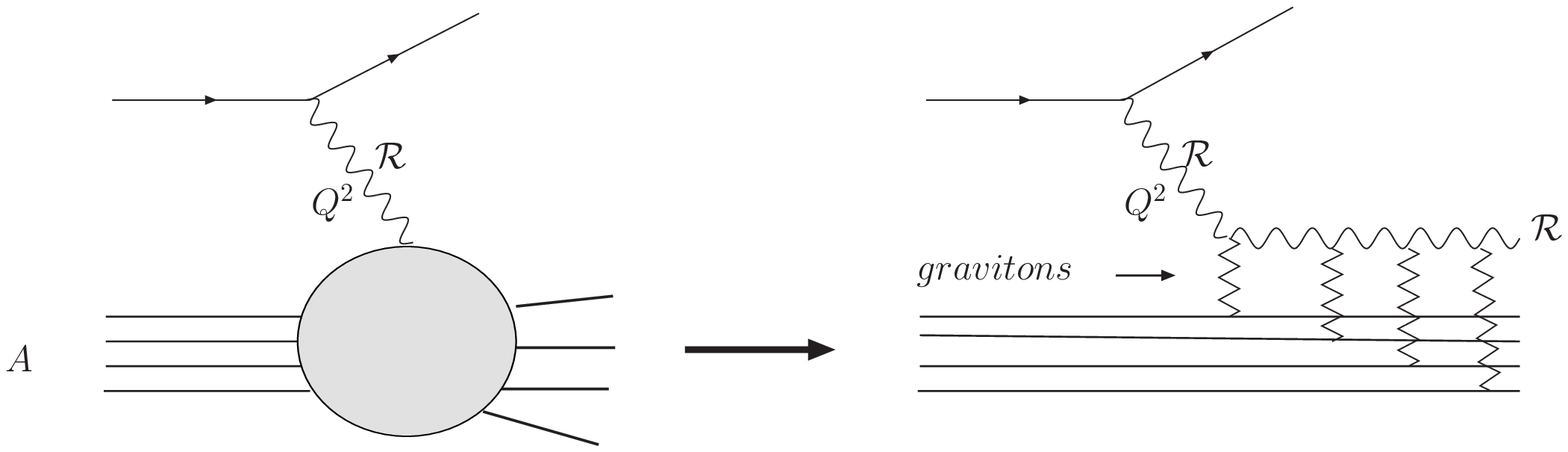,width=140mm,height=40mm}}
\caption{DIS with the nuclear target. The wave line denotes the ${\cal R}$ current (${\cal R}$-boson),  while the zigzag lines show the graviton exchanges. $Q^2$ is the virtuality of the probe.}
\label{disn} }

The solution to \eq{DIS1} is

\beq \label{DIS2}
 \Psi_{\cal R}\Lb Q^2,z\Rb \,\,=\,\,Q\,z K_1\Lb Q\,z\Rb
\eeq

However, ${\cal R}$ - boson is a vector with $d +1$ components. The careful analysis of ref. \cite{POST} shows that \eq{DIS2} describes only $d$ components of this vector, while the $ (d + 1) $-th component has a different dependence on $Qz$. Finally\cite{MHI},

\beq \label{DIS3}
|\Psi\Lb Q^2, z \Rb|^2\,\,=\,\,\Lb K^2_1\Lb Q\,z\Rb\,\,+\,\,K^2_0\Lb Q\,z\Rb\Rb\,z^3
\eeq

The deep inelastic structure function has the following form\cite{POST,MHI}

\bea \label{DIS4}
&&F_2\Lb Q^2, x= Q^2/s\Rb \,\,=\\
&&\,\,\,\,\,\,\,\,\,\,\,\,C\,\alpha'\,Q^6\,\int\,d^2 b \int\,d z \,z^3\,\Lb K^2_1\Lb Q\,z\Rb\,\,+\,\,K^2_0\Lb Q\,z\Rb\Rb\,\,2\,\,Re\left\{ 1\,\,-\,\,\exp\Lb  i\,\frac{ g^2_0\,N_c}{4}\,\,\frac{Q^2}{x}\,z^4\,S\Lb b\Rb\Rb\right\}\nonumber
\eea

where $C$ is a dimensionless constant.

Changing the variable $ z$ to $\zeta = Q\,z$, one can see that $F_2$ can be written in the form

\bea \label{DIS5}
&&F_2\Lb Q^2, x= Q^2/s\Rb \,\,=\,\,\,C\,\,Q^2 \,\int\,d^2 b\,\Phi\Lb \tau(Q,x,b)\Rb\,=\\
&&\,\,\,\,\,\,\,\,\,\,\,\,=C\,\,Q^2\,\int\,d^2b\,\int\,d \zeta \,\zeta^3
\Lb K^2_1\Lb \zeta\Rb\,\,+\,\,K^2_0\Lb \zeta \Rb\Rb
\,\,\,Re\left\{ 1\,\,-\,\,\exp\Lb  i\,\frac{ 1}{\tau}\,\zeta^4\Rb\right\}\nonumber
\eea

where

\beq \label{TAU}
\tau\,\,=\,\,\frac{Q^2\,x}{\frac{g_0^2 N_c}{4} \,S\Lb b \Rb}\,\,=\,\,\frac{Q^2}{Q^2_s}
\eeq

One can see that the DIS structure function shows the geometrical scaling behavior with the saturation momentum, which we can find from the equation with $\tau =1$ . It is equal to

\beq \label{QS}
Q^2_s\Lb x \Rb \,\,=\,\,\,g^2_0\,N_c\,S\Lb b\Rb/(4\,x)\,\,\,\propto\,\,\frac{A^{\frac{1}{3}}}{N_c}\,\frac{1}{x}
\eeq

Therefore, $F_2$ shows the same main features as $F_2$ in high density QCD \cite{GLR,MUQI,MV,GESC}, namely
the geometrical scaling behavior, large values of the saturation scale in the region of low $x$, and the expected  dependence of $Q^2_s \propto \,A^{1/3}$. Actually, our analysis of $Q_s$ repeats the one in Ref. \cite{MHI}, and the difference between them stems from our integration over the impact parameters.

One can see from \fig{gsclf} that the function $\Phi$ has the same
behavior as we expected from high density QCD, namely it approaches
unity at small values of $\tau$.  Such a behavior  looks strange,
especially if we compare this function with \eq{GLF3}, which leads
to an amplitude that oscillates between 0 and 2. Let us consider
$\tau \,>\,1$. In this case, we can replace the modified Bessel
functions (McDonald functions) in \eq{DIS5}  by their asymptotic
expression, namely, $K_n(\zeta) \,\xrightarrow\,\sqrt{2
\pi/\zeta}\,\exp( - \zeta)$, and in this case \eq{DIS5} has the form

\bea \label{DIS51}
&&F_2\Lb Q^2, x= Q^2/s\Rb \,\,=\,\,\,C\,\,Q^2 \,\int\,d^2 b\,\Phi\Lb \tau(Q,x,b)\Rb\,=\\
&&\,\,\,\,\,\,\,\,\,\,\,\,=C\,\,2\,\pi\,Q^2\,\int\,d^2 b\,\int\,d \zeta \,\zeta^2\,e^{ - \zeta}\,
\,\,\,Re\left\{ 1\,\,-\,\,\exp\Lb  i\,\frac{ 1}{\tau}\,\zeta^4\Rb\right\}\nonumber
\eea

The second term in $\{\dots \}$ can be estimated by the saddle point method. One can see that the saddle point value for $\zeta = \zeta_{SP} =\,\Lb - i \tau/3\Rb^{1/3}$, and the integral has the following
 form

\beq \label{DIS52}
\Phi_{SP}(\tau) \,=\,1 \,-\,\sqrt{\frac{\pi\,\tau}{12 z^2_{SP}}}\,z^2_{SP}\,e^{-(2/3)(i \tau/3)^{1/3}}\,\,\xrightarrow\,\,1
\eeq

One can see that at $\tau \to 0$, the exponent $e^{-(2/3)(i \tau/3)^{1/3}} \,\to\,1$, but the pre-exponential factor $\propto \tau^{5/6} $ vanishes. However, since $\zeta_{SP} \ll 1$, at small values of $\tau$ we have to use the expression for the modified Bessel function at $\zeta \to 0$, namely $K_n(\zeta)
\xrightarrow{\zeta \to 0} 1/\zeta^n$. Doing this, one can see that $\zeta \sim \tau^{1/4}$ contributes to the integral  leading to the behavior of the second term in \eq{DIS52} proportional to $\sqrt{\tau}$.

The above discussion shows that predictions of high density QCD differ from  those of N=4 SYM, only in the way that $\Phi(\tau)$ approaches unity, namely $\Phi - 1 \propto \exp\Lb - C \ln^2(1/\tau)\Rb$ in high density QCD, and $   \Phi - 1 \propto \exp\Lb - \h \ln(1/\tau)\Rb$, in our approach.

We need to integrate $ \Phi\Lb \tau(b)\Rb$ over $b$ (see \eq{DIS5}), to obtain the total cross section
for DIS

\bea \label{DIS6}
\sigma_{tot}\Lb DIS\Rb\,\,&=&\,\,\frac{4 \pi^2}{Q^2} \,F_2\Lb Q^2, x= Q^2/s\Rb \,\\
        &=&\,C\,\int\,d^2 b \,\Phi\Lb \tau(b)\Rb \,=\, 2 \pi\, C \int^\infty_{\tau(b=0)}
\frac{d \tau}{\tau} \frac{S\Lb b(\tau)\Rb}{S_{b^2}
\Lb b(\tau) \Rb}\,\Phi\Lb \tau\Rb   \nonumber \\
    &\xrightarrow{x \rightarrow 0}&\,C\,\pi  R^2_A\,
\int ^\infty_{\tau(b=0)\leq \,\tau_{max}} \,d \tau \,\,\frac{\Phi\Lb\tau\Rb}{\tau}\,\,R\Lb \tau\Rb
\,\,\mbox{where}\,\,\,R\Lb \tau\Rb \,\,=\,\, \frac{S\Lb b(\tau)\Rb}{S_{b^2}\Lb b(\tau) \Rb}
\nonumber
\eea
where $\tau = \tau_{max} $ is the position of the maximum of the function $\Phi(\tau)$.
The explicit form of the function $R(\tau)$ depends on the  dependence of $S(b)$ on the impact parameter.
We list below this function for several nucleus models:

\beq \label{MOD}
R\Lb \tau \Rb\,\,=\,\,\left\{ \begin{array}{c l l}\,\tau &\mbox{cylindrical nucleus} &  S(b) =(A/\pi R^2_A)\,\Theta\Lb R_A - b\Rb;\\
 ~\\ 1 & \mbox{Gaussian form} & S\Lb b \Rb = (A/\pi R^2_A)\,\exp\Lb - b^2/R^2_A\Rb; \\
~\\
\tau(b=0)/\tau^3\,\,\,\,\, & \mbox{spherical drop nucleus}\,\, \,\,\,& S\Lb b \Rb = (3 A/4\pi^2\,R^2_A)\,\sqrt{R^2_A - b^2};
\end{array} \right.
\eeq

\FIGURE[h]{\begin{minipage}{90mm}
\centerline{\epsfig{file=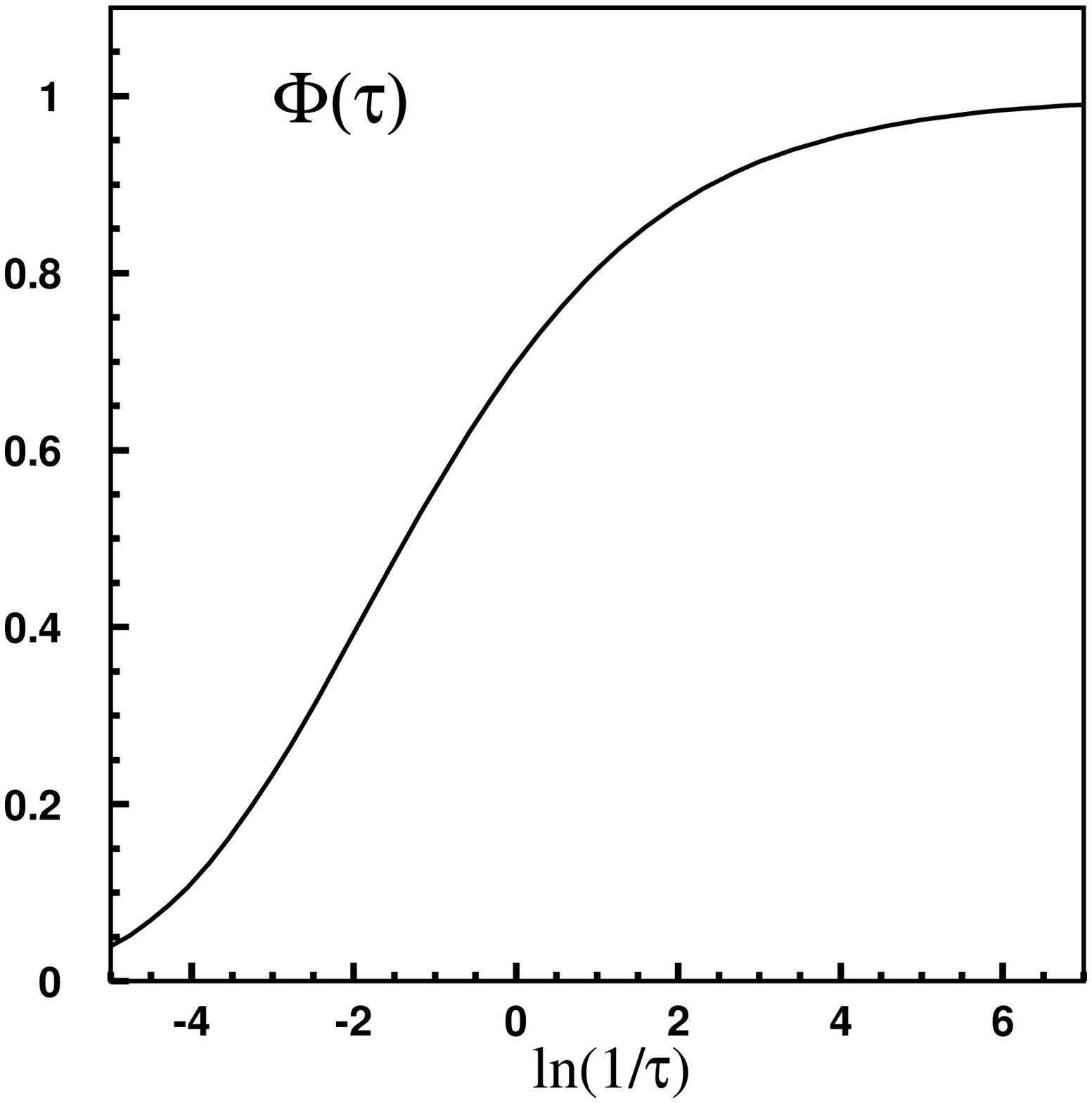,width=80mm}}
\caption{The $\tau$  dependence of  function $\Phi$.  }
\end{minipage}
\label{gsclf}}
Unfortunately, in a realistic model of the nucleus with the
Wood-Saxon form for the $b$ dependence, we cannot give a simple
analytical form of the function $R(\tau)$.  In \fig{sigtf} we plot
the integral over $\tau$ in \eq{DIS6}, for Gaussian $b$
distribution. This distribution, being oversimplified, leads to
correct estimates for the average characteristics of nuclei.

>From \eq{DIS6}, one can see that the total cross section for DIS will be $2 \pi R^2_A \times \ln(1/\tau(b=0))$, once more in  accordance with our expectation from high density QCD for such $S(b)$.
In the case of the Wood-Saxon  parameterization,\,\,\,\,\, $S\Lb b \Rb \xrightarrow{ b > R_A} \exp\Lb - b/h \Rb $
which leads to $\sigma_{tot} \,\propto \,\ln^2(\tau(b=0)$.  This behavior coincides with the expectation of high density QCD.

Therefore, the Glauber-Gribov approach leads to a behavior of the DIS structure function, which fully supports the high density QCD picture, reproducing the geometrical scaling behavior, and the existence of only one new scale, namely the saturation momentum.

The main difference between N=4 SYM and high density QCD, lies only
in the relation between the total cross section and the cross
section of diffractive dissociation. That is, $\sigma_{tot}\Lb DIS
\Rb = \sigma_{diff}\Lb DIS \Rb$ for N=4  SYM, and $\sigma_{tot}\Lb
DIS \Rb \neq  \sigma_{diff}\Lb DIS \Rb$ but\\ $ \sigma_{diff}
\xrightarrow{x \to 0} \h \sigma_{tot}$ for high density QCD.   In
N=4 SYM, this equality means that the elastic cross section is equal
to zero, in sharp contradiction with QCD and any parton
interpretation of high energy scattering. However, this is a direct
consequence of the fact that the graviton has spin 2. Actually, it
has been shown in Ref.\cite{BST1} that its spin in N-4 SYM is not
exactly 2, but rather $j_{graviton} \equiv j_0\,=\, 2 -
2/\sqrt{\lambda}$. Because of this,  the amplitude of the
interaction with the nucleon is not purely real, as it is given by
\eq{N48}, but it has an imaginary part which is proportional to $2 -
j_0$.  \fig{totin} illustrates how this imaginary part influences
the total and inelastic cross sections.  We introduce the functions
$\Phi_{tot} $ and $\Phi_{in}$ as
$$ \sigma_{tot} \,=\,\int d^2 b\,\, \Phi_{tot}\Lb \tau \Rb\,\,\,\mbox{and}\,\,\,\,
\sigma_{in} \,=\,\int d^2 b\,\, \Phi_{in}\Lb \tau \Rb$$ The
functions $\Phi_{tot} $ and $\Phi_{in}$ are shown in \fig{totin},
for the imaginary part of the graviton exchange, which is $10\%$ of
the real part of the amplitude. One can see that such a small
imaginary part generates a large inelastic cross section, and
therefore the DIS structure function in N=4 SYM, with reggeized
graviton, leads to a qualitative picture which is very difficult to
differentiate from the high density QCD predictions.

\DOUBLEFIGURE[h]{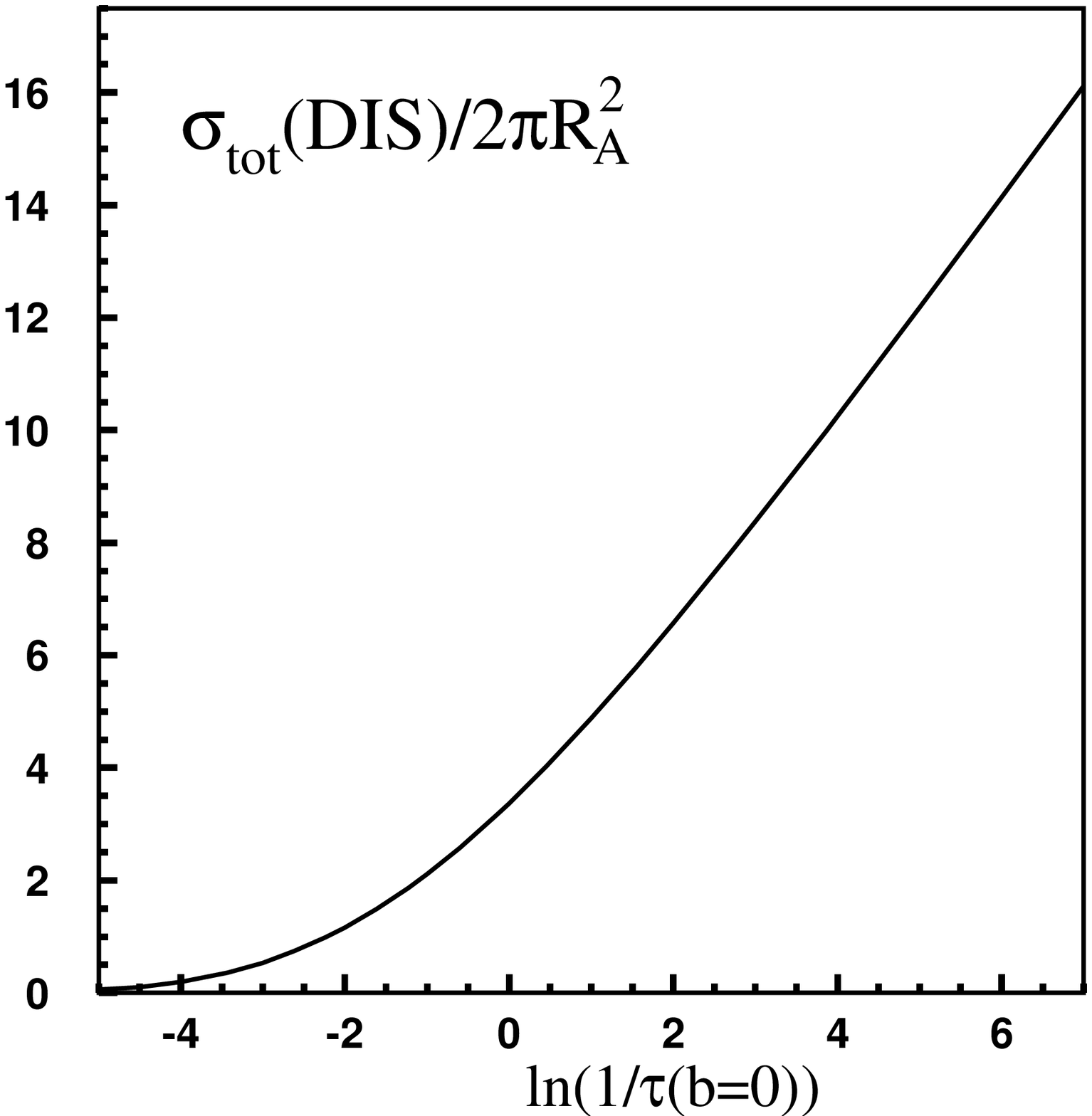,width=85mm,height=75mm}{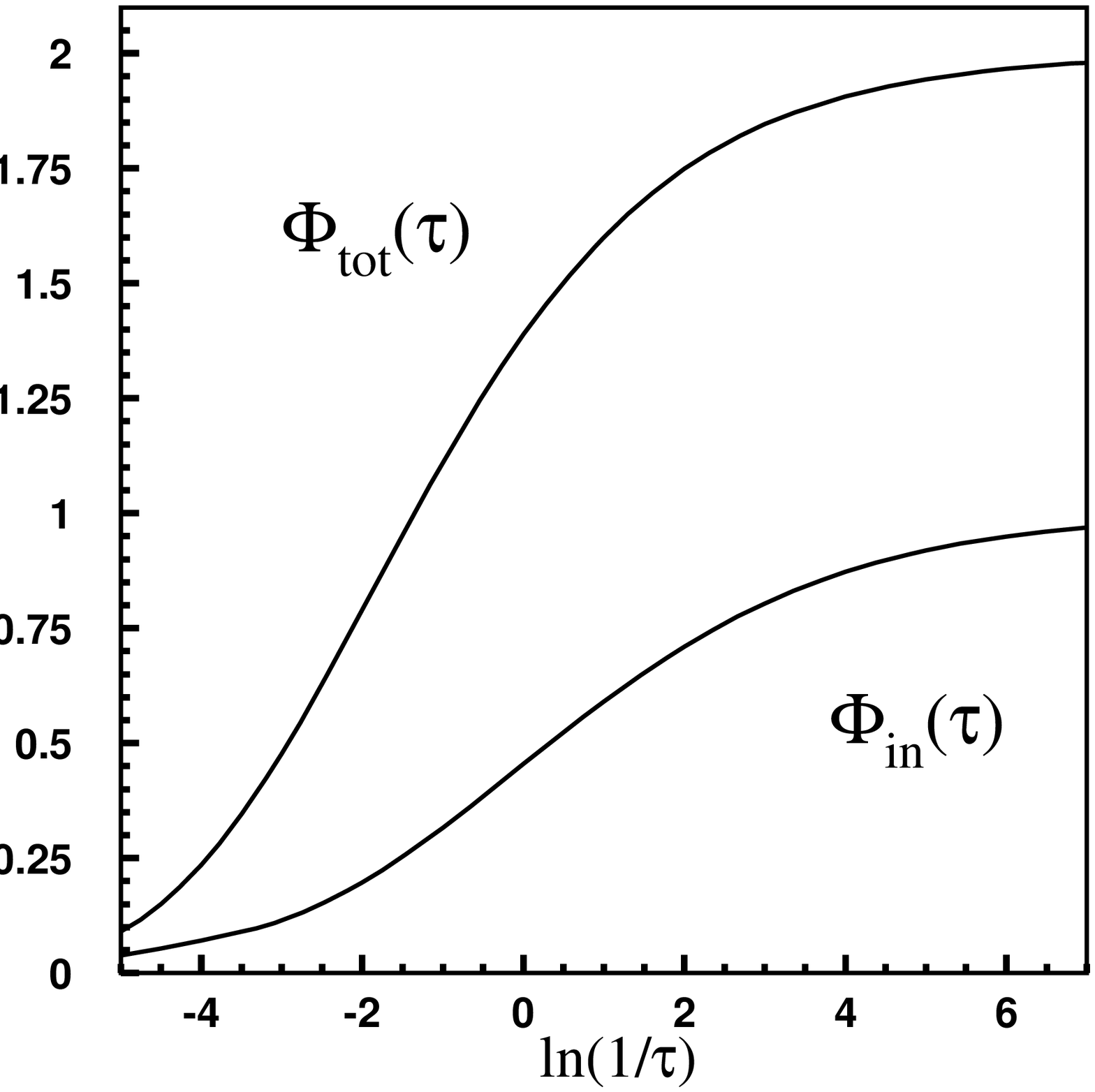,width=85mm,height=75mm}
{The integral over $\tau$ in \protect\eq{DIS6} as a function of $\tau(b=0)$ for Gaussian dependence of the  nucleon density in nucleus versus the impact parameter. \label{sigtf}}
{The behavior of the total and inelastic cross sections for the graviton exchange with $10 \%$ imaginary part of the amplitude.\label{totin}}

To complete the proof of \eq{DIS51},
we need to discuss the contributions from multi-graviton exchanges in the nucleon amplitude. At first sight, they should be essential, since each graviton exchange brings in a factor (see \eq{N48})

\beq \label{DIS7}
A^{G}_N\Lb s,b\Rb \,\,=\,\,i\,g_0 \,s\,\int d z' \,|\Phi(z')|^2\,z\,z'\,G_3(u)\,\,\xrightarrow{b\gg z' > z}\,\,\,
8\,i\,g_0\,s\, z^4 \,\int d z' \,|\Phi(z')|^2\,z'^4/\Lb b^2\Rb^3
\eeq

>From \eq{DIS7}, one can see that the amplitude $A^G_N(s,b) \gg 1$
for $b^2 = b^2_0 \propto \Lb  i s\,z^4\Rb^{1/3}$.  This means that
we need to take into account all terms of the order of  $\Lb
A^G_N\Rb^n$. Using the eikonal formula for summing such terms, we
see that  for the nucleon amplitude we have the following
expression, instead of the simple formula of \eq{N48},

\bea \label{DIS8}
 &&\int \,d^2\,b\,A_N\Lb s,b\Rb\,\,=\\
&&\int \,d^2 b \,\left\{ 1\,\,-\,\,\exp\Lb A^{G}_N\Lb s,b\Rb\Rb \right\}
\,\,=\,\,\int \,d^2 b \,\left\{ 1\,\,-\,\,\exp\Lb  i\,g_0 \,s\,\int d z' \,|\Phi(z')|^2\,z\,z'\,G_3(u)\Rb \right\} \nonumber
\eea

 The integral over $b$ can be estimated as $ \int \,d^2\,b\,A_N\Lb s,b\Rb\,\propto \,C\,\pi b^2_0 \propto
\Lb i \,s\,z^4\Rb^{1/3}$.
Using \eq{DIS8}, we can rewrite \eq{GLF1} in the form

\beq \label{DIS9}
A_A(s,b)\,\,=\,\,i\,\int \,d\,z \,|\Phi_p(z)|^2\,\left\{\,1\,\,-\,\,e^{ i\,\,C'\,\pi b^2_0  \,
S\Lb b\Rb} \right\}
\eeq

For DIS we have

\bea \label{DIS10}
&&F_2\Lb Q^2, x= Q^2/s\Rb \,\,=\\
&&\,\,\,\,\,\,\,\,\,\,\,\,C\,\alpha'\,Q^6\,\int\,d^2 b \int\,d z \,z^3\,\Lb K^2_1\Lb Q\,z\Rb\,\,+\,\,K^2_0\Lb Q\,z\Rb\Rb\,\,2\,\,Re\left\{ 1\,\,-\,\,\exp\Lb  i\,\pi C'\,\,\,b^2_0\,z^4\,S\Lb b\Rb\Rb\right\}\nonumber
\eea
where $C$ and $C'$ are dimensionless constants, whose values are irrelevant for our discussion.

Performing the integral over $z$ using the asymptotic behavior of modified Bessel  functions and the saddle point approach, one can see that the saddle point value of $z=z_{SP} $ is equal to

\beq \label{SP1}
z_{SP} =\frac{Q^3}{ i \,s\,S^3\Lb b\Rb} \,\approx\,\,\frac{Q^2}{i\, A\,s}
\eeq
where $A$ is the number of the nucleons in a nucleus.  The value of $F_2$ in the saddle point is

\beq \label{SP2}
F_2 \,\,\,\propto\,\,\, \exp\Lb -i \,const\,Q^4/(s A)\Rb
\eeq

This formula, if correct, leads to a saturation scale $Q^2_s \propto
A/x$, in drastic contradiction with the prediction of high density
QCD, both in the $A $ and $s$ dependencies. However, if we come back
to \eq{DIS7}, we obtain

\beq \label{SP3}
A^{G}_N\Lb s,b\Rb \,\,=\,\,\,
8\,i\,g_0\,s\, z^4 _{SP}\,\int d z' \,|\Phi(z')|^2\,z'^4/\Lb b^2\Rb^3\,\,\propto\,\,\,
i\, s\,\Lb \frac{Q^3}{A\,s}\Rb^4\,\,\ll \,\,1\,\,\mbox{for}\,\,\,\,s \gg\,\,Q^2 (x \to 0)
\eeq

>From these estimates we conclude that  the multi-graviton exchange does not contribute to DIS with a nuclear target, at low $x$.

\section{DIS with nuclei:  ultra high energy limit}

The result of the previous section is, however, valid only for a
limited range of energy. Indeed, we observe that the value of the
typical impact parameters in the  nucleon scattering amplitude (
$b^2= b^2_0 \propto \Lb  i s\,z^4\Rb^{1/3}$), grows with energy, and
for energies larger than the energy ($s = s_{crit}$)  when $b_0 \geq
R_A$, we cannot use the eikonal  formula in the form of \eq{GA14}.
Indeed, for such large energies, the main assumption of the
Glauber-Gribov approach does not work. This assumption has been
discussed in  \eq{GA8}, which can be rewritten in the following way
in the case of one graviton exchange

\beq \label{UHE1}
A_A\Lb s,b\Rb \,\,=\,\,\int\,d^2 b' \,A_N\Lb s,  b' \Rb\,S\Lb \vec{b} - \vec{b}'\Rb \,\rightarrow\,
\int\,d^2 b' \,A_N\Lb s,  b' \Rb\,S\Lb b\Rb
\eeq

In \eq{UHE1}, we assume that in the interaction with one nucleon, the typical impact parameters are much smaller than $R_A$, which gives the scale for the impact parameter distribution in the nuclei.
If the typical  $b$ in the nucleon interaction is larger than $R_A$, we have to use a different approximation, namely we need  to rewrite \eq{UHE1} in the form

\beq \label{UHE2}
A_A\Lb s,b\Rb \,\,=\,\,\int\,d^2 b' \,A_N\Lb s,  b' \Rb\,S\Lb \vec{b} - \vec{b}'\Rb \,\rightarrow\,
\,A_N\Lb s,  b \Rb\,\int\,d^2 b'\,S\Lb b\Rb \,\,=\,\,A\,\,A_N\Lb s,  b \Rb
\eeq
This equation leads to a new formula for the scattering amplitude with a nucleus, instead of \eq{GA14}, namely,

\beq \label{UNE3}
A_A\Lb s,b\Rb \,\,=\,\,i \Lb\,1\,\,\,-\,\,\,\exp \Lb i\, A\,  A_N(s, b)\Rb\Rb
\eeq
which leads to an expression for  the DIS structure function in the form

\bea \label{UHE4}
&&F_2\Lb Q^2, x= Q^2/s\Rb \,\,=\\
&&\,\,\,\,\,\,\,\,\,\,\,\,C\,\alpha'\,Q^6\,\int\,d^2 b \int\,d z \,z^3\,\Lb K^2_1\Lb Q\,z\Rb\,\,+\,\,K^2_0\Lb Q\,z\Rb\Rb\,\,2\,\,Re\left\{ 1\,\,-\,\,\exp\Lb  i\,A\,A_N\Lb s,b\Rb\Rb\right\}\nonumber
\eea
where $A_N\Lb s,b\Rb$ is given by \eq{DIS7}.  $A_N$ can be rewritten at large $b$, using \eq{GA3}  and \eq{DIS7}, in the form

\beq \label{UHE5}
A_N\Lb s,b;z\Rb\,\,\,=\,\,\,i\,\sin\left[ s\, \frac{\,g^2_0\,N_c}{8}\,\,\,z^4/(b^2)^3\, \right]\,\,+\,\,\,
2 \sin^2\left[s\, \frac{\,g^2_0\,N_c}{16}\,\,z^4/(b^2)^3\, \right]
\eeq

Substituting \eq{UHE5}, we do the integral over $z$ using the steepest decent method. The most important part of the nucleon amplitude is the imaginary part, which leads to a damping of the interaction matrix  ($S$-matrix) at high energies, provided the amplitude tends to unity.
The saddle point for $z$ is equal to

\beq \label{UHE6}
z_{SP}\,\,=\,\,\- b^2\,\,\Lb \frac{Q}{4 A s \cos\left[\frac{g^2_0\,N_c}{8}\,\,z^4/(b^2)^3\right]}\Rb^\frac{1}{3}
\eeq

Taking the integral using the steepest decent method we obtain

\beq \label{UHE7}
F_2\Lb Q^2, x= Q^2/s\Rb \,\,\,\propto\,\,Q^5\,\int\,d^2 b \,z^2_{SP} \,\frac{\sqrt{\pi}}{2 \,z_{SP}\sqrt{A\,s}}\,\exp \Lb - 5/4\,\,b^2\,\,\Lb - \frac{Q^4}{4 A\, s }\Rb^{\frac{1}{3}}\Rb
\eeq
where we replaced sines and cosines in \eq{UHE5} and \eq{UHE6}, by unity since these functions cannot change the energy and $Q$ dependence of the resulting amplitude.

>From \eq{UHE7}, one can see that the typical values of the impact parameters are large and equal to

\beq \label{UHE8}
b^2_0 \,\,=\,\,4/(5 Q z_{SP} )\,\,= \,\,\frac{4}{5}\,\Lb - \frac{4 A s}{Q^4}\Rb^{\frac{1}{3}}\,\,\gg\,\,z^2_{SP}
\eeq

The resulting answer for $F_2$ is the following

\beq \label{UHE9}
F_2 \,\,=\,\, \,\,\,\propto\,\,\alpha'\,Q^2\,\Lb \frac{A \,s}{Q^2}\Rb^{\frac{1}{3}} \,\,=\,\,
 \,\,\alpha'\,Q^2\,A^{\frac{1}{3}}\,x^{-\,\frac{1}{3}}\,
\eeq

Therefore, we see that we expect a very strange behavior from the
point of view of high density QCD, both as function of $A$ and $x$.
The origin is clear. N=4 SYM has a massless particle, namely the
graviton, and because of this the nucleon amplitude falls at large
$b^2 \gg z^2 + z'^2$,  as a power of $1/(b^2)^3$ . Such a power-like
behavior leads to a typical $b$ which grows as a  power of energy,
(see Ref.\cite{KOWI} for details), as has been demonstrated above.
However, as has been shown in Refs. \cite{BST1,BST3}, actually the
graviton has a mass which is not equal to zero if we dealing with
the propagation of the graviton in $AdS_5$.  This mass leads to a
reggeization of the graviton, which has spin $j_0 = 2 -
2/\sqrt{\lambda} \,<\,2$, in the scattering kinematic region where
the square of the momentum transferred  $t$ is negative ($t < 0$).
The fact that there is no massless particle in the curved space
means that at large $b$, the amplitude should falls exponentially
leading to a log  energy dependence of the cross section. This is
the reason why in the next section we will discuss the exchange of
the reggeized graviton, and the Glauber- type formula which such an
interaction induces.

\section{DIS with nuclei:  graviton reggeization.}

As has been discussed in Ref.\cite{BST3},  for the exchange of the
reggeized graviton we need to replace \eq{N48} by a more general
expression, namely

\beq \label{GR1}
\int\,d^2\,b\,A_N\Lb s, b \Rb\,\,=\,\,i\,g^2_0 \frac{1}{\tilde{s}}\,\left\{ - \int \,\frac{d\,j}{ 2 \pi \,i}\,\Lb \frac{ \tilde{s}^j \,+\,(- \tilde{s})^j}{\sin\,\pi j}\Rb \,\int\,d^2 b \,\,G_3\Lb u, j\Rb\right\}
\eeq
where

\bea \label{GR2}
G_3\Lb u, j\Rb \,\,&=&\,\,\frac{1}{4 \pi}\,\frac{\Lb 1 + u + \sqrt{u (u + 2)} \Rb^{2 - \Delta_+(j)}}{\sqrt{u (u + 2)}}\\
\mbox{with}\,& & \,\,\Delta_+(j)=2 \,+\,\sqrt{4 \,+\,2 \sqrt{\lambda} (j - 2)}\,=\,2 + \sqrt{2 \sqrt{\lambda} \,( j - j_0)} \nonumber
\eea

Using the definition of $u$ given in \eq{UTR} and \eq{DB}, we can easily evaluate the integral over $b$ in \eq{GR1}
with the following result

\beq \label{GR3}
\int\,d^2 b \,\,G_3\Lb u, j\Rb\,\,\,=\,\, z z'\,\,\frac{1}{2 - \Delta_+\Lb j \Rb}\,
\Lb 1 + u(b=0) \,+\,
\sqrt{u(b=0) \,\Lb u(b=0) + 2 \Rb} \Rb^{ 2 \,-\,\Delta_+\Lb j \Rb}
\eeq

The integral over $j$ in \eq{GR1} is a contour integral, and the contour is located to the right of all singularities of $\int\,d^2 b \,\,G_3\Lb u, j\Rb$, but to the left of $j =2$, and the contour can be drawn to be parallel to the imaginary axis.
In \eq{GR3}, one can see that our singularity in $j$  stems from the zero of
the factor  $2 - \Delta_+\Lb j \Rb$. Denoting $ \sqrt{2 \sqrt{\lambda} \,| j - j_0|} = \nu$, we can rewrite the contribution of the square root singularity at $j = j_0$ in the following way

\bea \label{GR4}
\int\,d^2\,b\,A_N\Lb s, b \Rb\,\,&=&\,\, \,g^2_0 \,2\,z\,z'\,\Lb  \cot \frac{\pi j_0 }{2}  \,+\,i \Rb
\,\,\tilde{s}^{j_0 - 1}\,\\
 & \times & \int^{i \epsilon + \infty}_{i \epsilon - \infty}\,\frac{d \nu}{ \sqrt{\lambda}\, \pi }\,\,
\exp\Lb - \nu^2/(2 \sqrt{\lambda}) + i \nu \ln\left\{1 + u(b=0) \,+\,
\sqrt{u(b=0) \,\Lb u(b=0) + 2 \Rb} \right\}\Rb \nonumber\\
 & \xrightarrow{z >z'} &\,\, \,g^2_0 \,2\,z\,z'\,\Lb  \cot \frac{\pi j_0 }{2}  \,+\,i \Rb
\,\,\tilde{s}^{j_0 - 1}\,\, \int^{i \epsilon + \infty}_{i \epsilon - \infty}\,\frac{d \,\nu}{   \sqrt{\lambda}\,\pi }\,\,
\exp\Lb - \nu^2/(2 \sqrt{\lambda}) + i \nu \,\ln\Lb \frac{z}{z'} \Rb \Rb\nonumber
\eea

In the course of deriving \eq{GR4}, we neglected in the signature factor the contribution of the term
 $  \nu^2/(2 \sqrt{\lambda})$, considering it to be smaller than $j_0$ ($ j_0 \,\gg \,\nu^2/(2 \sqrt{\lambda})$. The integral in \eq{GR4} can be evaluated such that it reduces to the following expression

\beq \label{GR5}
\int\,d^2\,b\,A_N\Lb s, b \Rb\,\,=\,\, \,g^2_0 \,2\,z\,z'\,\Lb  \cot \frac{\pi j_0 }{2}  \,+\,i \Rb
\,\,\tilde{s}^{j_0 - 1}\,\,\sqrt{\frac{2}{\pi\,\sqrt{\lambda}\,\,\ln \tilde{s}}}\,\exp \Lb - \frac{\sqrt{\lambda}\,\ln^2 (z'/z)}{2 \ln \tilde{s}}\Rb
\eeq

The result of \eq{GR5} is obtained assuming that $\lambda$ is fixed, but $s \to \infty$. From \eq{GR1}, \eq{GR2} and \eq{GR3}, we can recover a different limit, namely $\lambda \,\to \,\infty$ when $ \tilde{s} \,\gg\,1$. Indeed, in this limit $\Delta_{+} \to 4 + ( \sqrt{\lambda}/2) \,( j - 2)$.
Since $ 2 - \Delta_{+}(j)\, \to \, ( \sqrt{\lambda}/2) \,( j - 2)$, we can close the contour on the pole
which stems from $ 2 - \Delta_{+}(j) = 0$.  The signature factor can be rewritten in the form

\beq \label{GR6}
\Lb  \cot \frac{\pi j_0 }{2}  \,+\,i \Rb\,\,\xrightarrow{ \lambda \to \infty} \,\,\frac{\sqrt{\lambda}}{\pi}
\eeq

Collecting everything together we obtain

\beq \label{GR7}
\int\,d^2\,b\,A_N\Lb s, b \Rb\,\,=\,\, \,g^2_0 \,2\,z\,z'\,\left\{ 1 \,\,-\,\,i \frac{2}{\sqrt{\lambda}} \right\}\,\,\int\,d^2\,b\,\,G_3\Lb u,2\Rb \,\,\,\equiv\,\, \,g^2_0 \,2\,z\,z'\,\left\{ 1 \,\,-\,\,i \frac{2}{\sqrt{\lambda}} \right\}\,\,\int\,d^2\,b\,\,G_3\Lb u\Rb
\eeq

We have used \eq{GR7} in our estimates of the value of the imaginary part of the nucleon scattering amplitude. \eq{GR7} leads to the exchange of the graviton with a small imaginary part, and this case has been considered in detail in this paper.\\

We concentrate our efforts on the limit $\tilde{s} \to
\infty$,$ \lambda = Const$. For this region we need to use
\eq{GR4} for the nucleon amplitude.  However, even more important
for the high energy behavior of the amplitude, is the fact that the
graviton has a mass in curved space (see \fig{pomn4}). Therefore,
the  graviton trajectory which gives the dependence of the spin of
the graviton on its mass,  has the intercept $j_0 =
\alpha_{graviton}(0) = 2 - 2/\sqrt{\lambda}$, and the mass of the
graviton is equal to
$m^2_{graviton}2/(\sqrt{\lambda}\,\alpha'_{graviton})$\footnote{In
the previous sections we called $\alpha'_{graviton}$ just
$\alpha'$.}.  Therefore, in N=4 SYM  all particles

\FIGURE[ht]{\begin{minipage}{90mm}
\centerline{\epsfig{file=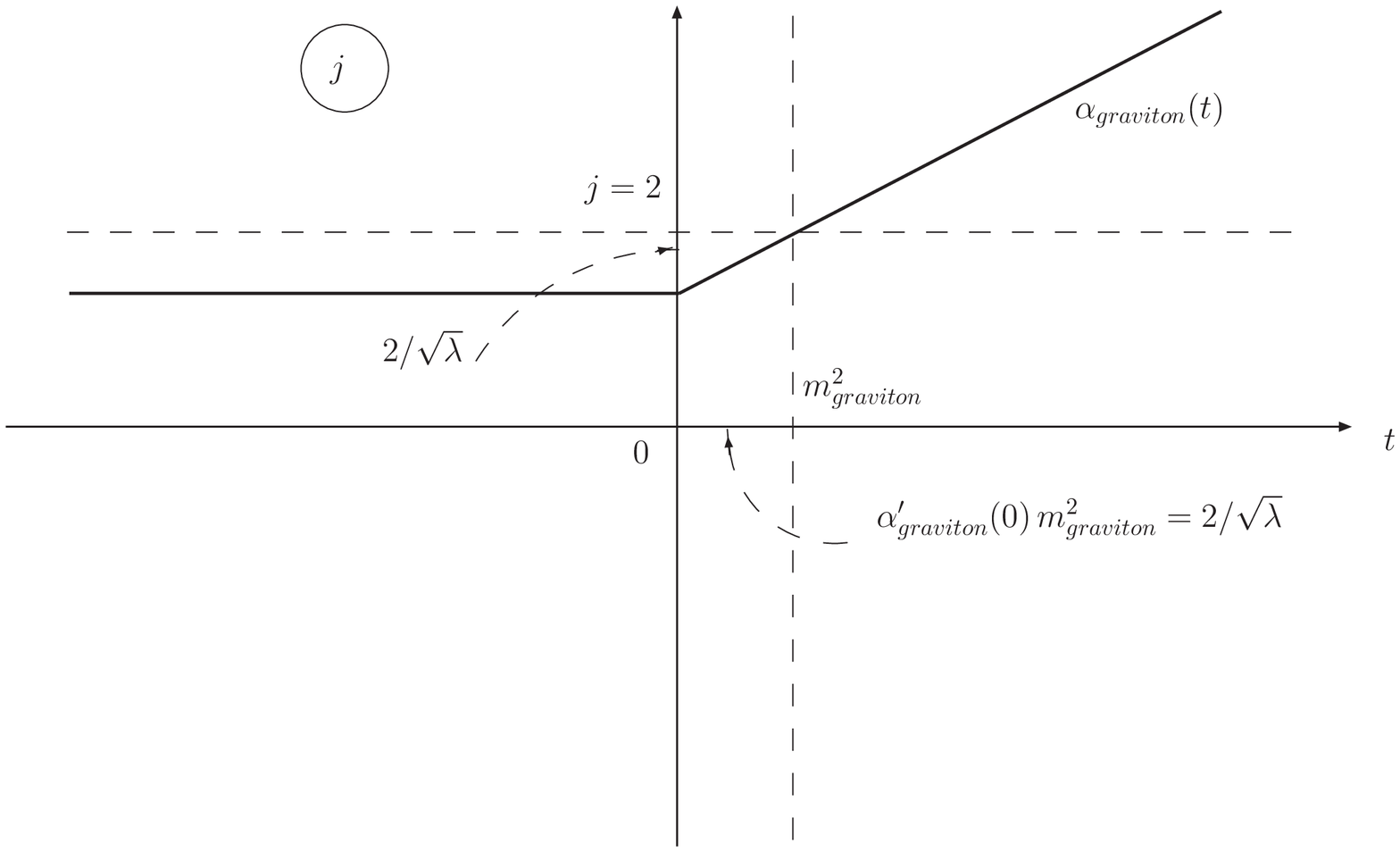,width=80mm}}
\caption{The  graviton(Pomeron) trajectory in N=4 SYM as it follows from Ref. \cite{BST1}.  }
\end{minipage}
\label{pomn4}}

 have masses, and the graviton is the lightest one.  In such a theory, the large $b$ dependence is determined by the mass of the lightest particle \cite{FROI}, namely $A_N(s,b) \,\,\to\,\, \exp\Lb - m_{graviton} \,b \Rb$.
This fact changes completely the ultra high energy behavior, that
has been considered in the previous section.  Assuming that the
graviton mass is small, we can distinguish four different kinematic
ranges of energy in the case if $R_A\,<\,1/m_{graviton}$; $z^2\,
g_0^2 s \,\leq\,1$; $z^2 g_0^2 s \,\geq\,1$, but $ b^2_0  \propto
\,z^2 s \,\leq\,R^2_A$; and $R^2_A \,\leq b^2_0  \propto \,z^2
s\,\leq\,1/m^2_{graviton}$ and $ b^2_0  \propto \,z^2 s\,$
$\geq\,\,1/m^2_{graviton}$. Nevertheless, we believe that  the mass
of the graviton should be such that $R_A \gg$ $ 1/m_{graviton}$ if
N=4 SYM pretends to describe the main features of the strong
interaction. Indeed, we know experimentally that the lightest hadron
is the $\pi$ meson, and the large $b$ dependence of the amplitude is
proportional to $\exp\Lb - b/2m_{\pi}\Rb$.  For a massive graviton
the amplitude falls as $\exp\Lb - b/m_{graviton} \Rb$. Therefore, to
avoid contradiction with experiment, we need to assume that
$m_{graviton}
> 2 m_{\pi}$. Having this in mind, we will consider a modification
to our formulae of the previous sections for the Glauber - Gribov
approach in the case of the reggeized graviton, in three kinematic
regions, which are $ z^2\,g_0^2 \,s \,\,\leq\,\,1$, where we can
restrict ourselves to the exchange of one graviton in the nucleon
scattering amplitude; $ z^2\,g_0^2 \,s \,\,\geq\,\,1 $ but $b^2_0
\propto \,z^2 s \,\leq\,1/m_{graviton} \leq\,R^2_A$ ( in this region
the multi graviton exchange could be essential); and the asymptotic
region where  $ z^2\,g_0^2 \,s \,\,\geq\,\,1 $ but $b^2_0 \propto
\,z^2 \,s\,\geq\,1/m_{graviton} \leq R_A$. Of course, we can
consider the kinematic region where $(1/m_{graviton}\,\ln s) \geq
R_A$, but in this region nuclei behave in the same way as the
nucleons, and we are not interested in this region.

\begin{boldmath}
\subsection{ $ z^2\,g_0^2 \,s \,\,\leq\,\,1$}
\end{boldmath}

In this kinematic region we can restrict ourselves to one reggeized graviton (Pomeron)  exchange,
and use \eq{GR5} instead of \eq{N48}. Therefore, we have

\bea \label{GR8}
&&F_2\Lb Q^2, x= Q^2/s\Rb \,\,=
\,\,\,\,\,C\,\alpha'\,Q^6\,\int\,d^2 b \int\,d z \,z^3\,\Lb K^2_1\Lb Q\,z\Rb\,\,+\,\,K^2_0\Lb  Q\,z\Rb\Rb\,\,\,2\,\,Re\left\{ 1\,\,-\right. \\
&&\left.-\,\,\exp\Lb  i\,\,g^2_0 N_c \int\,d z'\,| \Phi(z')|^2\,2\,z\,z'\,\Lb  \cot \frac{\pi j_0 }{2}  \,+\,i \Rb
\,\,\tilde{s}^{j_0 - 1}\,\,\sqrt{\frac{2}{\pi\,\sqrt{\lambda}\,\,\ln \tilde{s}}}\,\exp \Lb - \frac{\sqrt{\lambda}\,\ln^2 (z'/z)}{2 \ln \tilde{s}}\Rb \,S\Lb b\Rb\Rb\right\}\nonumber
\eea

Two features of \eq{GR8} are quite different from \eq{DIS4}, namely
that the nucleon amplitude has an imaginary part and shows a
different $z$ dependence. Roughly speaking, in \eq{GR8}, $A_N
\propto z^2$ instead of  $A_N \propto z^4$ in \eq{DIS4}.  The
integral over $z$ in \eq{GR8} can be evaluated using the steepest
descent method.  Using the asymptotic expression for  the modified
Bessel  function, we reduce \eq{GR8} to the following expression

\bea \label{GR9}
  &&F_2\Lb Q^2, x= Q^2/s\Rb \,\,= \\
&&\,\,\,\,\,\,\,\,\,\,\,\,\,\,\,\,\,C\,\alpha'\,Q^5\,\int\,d^2 b \int\, z^2\,d z\,e^{ - Q\,z}\,\,Re \left\{ 1 - \exp\Lb i g^2_0 \,N_c\,\xi(j_0)
\,s^{j_0 - 1}\, (z\,z')^{j_0}\,S\Lb b \Rb\,E\Lb \ln(z'.z) \Rb \Rb \right\} \nonumber\\
&& \mbox{where}\,\,\,\,\,E\Lb \ln(z'.z) \Rb\,\,=\,\,\sqrt{\frac{2}{\pi\,\sqrt{\lambda}\,\,\ln \tilde{s}}}\,\exp \Lb - \frac{\sqrt{\lambda}\,\ln^2 (z'/z)}{2 \ln \tilde{s}}\Rb \,\,\,\,\,\mbox{and}\,\,\,\,\,\xi(j) =
i\,\cot \frac{\pi j_0 }{2}  \,-\,1 \nonumber
\eea

Actually in \eq{GR9}, we need to integrate over $z'$, but we assume that the typical $z' \approx 1/\Lambda$, where $\Lambda$ is a scale of hadrons (glueballs) in N=4 SYM, and we can replace it with some average value.

It is convenient to introduce new dimensionless variables $\hat{Q} = Q \,z'$, $ \hat{s} = s z'^2$,$\hat{S}\Lb b \Rb = z'^2 S\Lb b \Rb$, $\hat{z}= z/z'$, for which
the equation for the saddle point  reads as follows

\beq \label{GR10}
\hat{Q}\,\,=\,\,g^2_0\,N_c\,\xi(j_0)\,\hat{S}\Lb b \Rb \,\hat{s}^{j_0 -1}\,\hat{z}^{j_0 - 1}_{SP}\,\Lb j_0\, - \frac{\sqrt{\lambda}\,\ln(1/\hat{z}_{SP})}{ \ln \hat{s} + \ln\hat{z}_{SP}}\,\Rb E\Lb \ln(1/\hat{z}_{SP})\Rb
\eeq
Rewriting \eq{GR10} in the form

\beq \label{GR11}
\ln \Lb \frac{\hat{Q}}{\hat{S}\Lb b \Rb}\Rb = (j_0 - 1) \,t \,-\,\frac{\sqrt{\lambda}\,( t - \ln\hat{s})^2}{2\,t} \,+\,w\Lb \hat{z}_{SP} \Rb
\eeq
where $w$ is a smooth function of $\hat{z}_{SP}$, and $t = \ln\Lb \hat{s}\,z_{SP}\Rb$. The approximate solution for $t $ is

\bea \label{GR12}
t^{\pm}\,\,&=&\,\,\ln \hat{s} - \frac{1}{\sqrt{\lambda}}\ln\Lb \frac{\hat{Q}}{\hat{S}\Lb b\Rb}\,e^{-w(0)}\Rb\,\,\pm\,\sqrt{\frac{-1 + j_0}{\sqrt{\lambda}}}\,\ln \hat{s}\,;\,\,\,\, \nonumber \\
 \hat{z}^{\pm}_{SP}\,\,&=&\,\,\exp\Lb - \frac{1}{\sqrt{\lambda}}\ln\Lb \frac{\hat{Q}}{\hat{S}\Lb b\Rb}\,e^{-w(0)}\Rb\,\,\pm\,\sqrt{\frac{- 1 + j_0}{\sqrt{\lambda}}}\,\ln \hat{s}\Rb\,;
\eea

Using \eq{GR12}, we find that the DIS structure function is proportional to

\bea \label{GR13}
F_2\Lb Q^2,x\Rb \,\,&\propto &\,\, Q^5\,\exp \Lb - \frac{j_0 -1}{j_0}\,\hat{Q}\,\hat{z}_{SP}\Rb\,\,
=\,\,Q^5\,\exp \Lb -\Lb Q/Q_s(x;A)\Rb^{ 1 - \frac{1}{\sqrt{\lambda}} -\sqrt{\frac{ j_0 - 1}{\sqrt{\lambda}}}}\Rb \\
&=&\,\,Q^5 \exp\Lb - \frac{j_0 -1}{j_0}\,\hat{Q}\,\left\{
\frac{Q}{g^2_0 N_c \xi(j_0)\,\hat{S}\Lb b \Rb }\right\}^{-\frac{1}{\sqrt{\lambda}}}\,\times\,
\hat{s}^{-\sqrt{\frac{ j_0 - 1}{\sqrt{\lambda}}}} \,\times\,
\Lb \frac{ j_0 - \frac{\sqrt{\lambda}\,\ln(1/\hat{z}_{SP})}{ \ln \hat{s} + \ln \hat{z}_{SP}}}{\sqrt{\frac{
2}{\pi \,\sqrt{\lambda} (\ln \hat{s} + \ln \hat{z}_{SP})}}}\Rb^{-\frac{1}{\sqrt{\lambda}}}\Rb\nonumber
\eea

In \eq{GR13} we chose $z^{-}_{SP}$, since it gives a larger contribution.
The saturation momentum is equal to

\beq \label{GR14}
Q_s\Lb x;A\Rb\,\,=\,\,\left\{ \frac{j_0}{j_0  -1}\,\Lb \frac{1}{x}\Rb^{\sqrt{\frac{j_0 - 1}{\sqrt{\lambda}}}}\Lb g^2_0 N_c \xi(j_0)\,\hat{S}\Lb b \Rb
\Rb^{\frac{1}{\sqrt{\lambda}}}\, \,\Lb \frac{ j_0 - \frac{\sqrt{\lambda}\,\ln(1/\hat{z}_{SP})}{ \ln \hat{s} + \ln \hat{z}_{SP}}}{\sqrt{\frac{
2}{\pi \,\sqrt{\lambda} (\ln \hat{s} + \ln \hat{z}_{SP})}}}\Rb^{-\frac{1}{\sqrt{\lambda}}}\right\}^{\frac{1}{j(\lambda)}}
\eeq
where

$$\mbox{with}\,\,\,j(\lambda) \,=\,1 - \frac{1}{\sqrt{\lambda}} -\sqrt{\frac{ j_0 - 1}{\sqrt{\lambda}}} $$
>From \eq{GR14} one can see that $F_2$ has a geometrical scaling behavior, if we neglect the log dependence of the saturation scale. The most interesting result is the fact that $Q_S \propto \,\Lb S\Lb b \Rb\Rb^{\frac{1}{\sqrt{\lambda}\,j(\lambda)}}\,\Lb 1/x\Rb^{\sqrt{\frac{j_0 -1}{\sqrt{\lambda}}}\,\frac{1}{j(\lambda)}}$.
At very large $\lambda$, the saturation momentum is constant and does not depend on $A$ and $x$. However, the $A$ dependance is more suppressed, namely $A^{1/(3 \sqrt{\lambda})}$, while the $x$ dependence has a suppression, which is however a much weaker one $(1/x)^{\lambda^{-1/4}}$. Such a behavior is similar to  what we expect in high density QCD for the running QCD coupling \cite{AQS}.

\begin{boldmath}
\subsection{ $ z^2\,g_0^2 \,s \,\,\geq\,\,1$ but $b^2_0  \propto \,z^2\,s \,\leq\,1/m_{graviton} \leq\,R^2_A$ }
\end{boldmath}

In this kinematic region we have to take into account the
multi-graviton interaction in the nucleon scattering amplitude.  At
high energy, the exchange of one Pomeron (reggeized graviton)
induces an imaginary part of the amplitude, as has been discussed,
which increases with energy. Such an increase leads to a nucleon
cross section of the order of $2 \pi b^2_0(x)$, where $b_0$ can be
estimated using the following equation

\beq \label{GR15}
A^G_N (s,z,b_0)\,\,\approx\,\,1/2
\eeq

The nucleon amplitude for single reggeized graviton exchange can be
evaluated using \eq{GR2} and the fact that $u(b) \to b^2/(2 z z')$
at large $b$. Repeating the same procedure, we obtain that (with
$\hat{b} = b/z'$)

\beq \label{GR16}
 A^G_N (s,z,b)\,\,\xrightarrow{b^2 \geq z^2  < z'^2}\,\,\frac{\hat{z}}{\hat{b}^2} \hat(s)^{j_0-1} \,\hat{z}^{j_0-1}\,\,\exp\Lb - \frac{\sqrt{\lambda}\,\ln^2 \Lb \frac{2 \hat{z}}{\hat{b}^2}\Rb}{2( \ln \hat{s} + \ln \hat{z})}\Rb
\eeq

>From \eq{GR15} and \eq{GR16}, we see that $\hat{b}^2_0\,\approx\,\hat{s}\,\hat{z}^2$. Therefore,

\beq \label{GR17}
F_2 \,\,\,\propto\,\,\exp\Lb - \hat{Q}\hat{z}  - Const\,S\Lb b \Rb\,\hat{s}^{j_0-1}\,\hat{z}^{j_0}\Rb
\eeq

which leads to small values of the typical $\hat{z} = \hat{z}_{SP}$, namely,

\beq \label{GR18}
\hat{z}_{SP} \,\,=\,\,\frac{\Lb\hat{Q/S\Lb b \Rb}\Rb^{\frac{1}{j_0 -1}}}{j_0\,\hat{s}}
 \eeq

At high energies, $z_{SP}$ is small, and the nucleon amplitude turns
out to be small even at small values of $b$. Therefore, we do not
need to discuss this region separately, and the answer is just the
same as in the previous section.

\begin{boldmath}
\subsection{ $ z^2\,g_0^2 \,s \,\,\geq\,\,1$ but $b^2_0  \propto \,z^2\,s \,\geq\,1/m_{graviton} \leq\,R^2_A$ }
\end{boldmath}

At such large impact parameters, we cannot use \eq{GR1} and \eq{GR5}. The main contribution in this region stems from the exchange of the lightest hadron (in our case of the graviton) \cite{FROI}, which has the form given in \eq{N41}, and can be written as

\beq \label{GR19}
A(s,b \gg z') \,\,\propto \,\,i\,g^2_0 \,s\,z^4\,\exp\Lb - m_{graviton} \,b\Rb
\eeq

The typical impact parameter can be found from the equation $ A\Lb
\eq{GR19}; s , b \Rb \approx 1/2$, which gives $b_0
\,=\,(1/m_{graviton})\,\ln \Lb g^2_0\, z^4 \,s\,\Rb$. Therefore for
$F_2$ we have

\bea \label{GR20}
  &&F_2\Lb Q^2, x= Q^2/s\Rb \,\,= \\
&&\,\,\,\,\,\,\,\,\,\,\,\,\,\,\,\,\,C\,\alpha'\,Q^5\,\int\,d^2 b \int\, z^2\,d z\,e^{ - Q\,z}\,\,Re \left\{ 1 - \exp\Lb i g^2_0 \,N_c\,\xi(j_0)
\,2\,\pi\,S\Lb b \Rb \,(1/m^2_{graviton})\,\ln^2 \Lb g^2_0\, z^4 \,s\,\Rb \Rb \right\} \nonumber
\eea

In \eq{GR20}, the main contribution stems from $z \propto \,1/Q$, which leads to

\bea \label{GR21}
&&F_2\Lb Q^2, x= Q^2/s\Rb \,\,= \\
&&\,\,\, \,\,\, \,\,\, \,\,\, C \,\alpha'\,Q^2\,\int\,d^2 b\,\,Re
\left\{ 1\,\,\,-\,\,\exp\Lb i g^2_0 \,N_c\,\xi(j_0) \,2\,\pi\,S\Lb b
\Rb \,(1/m^2_{graviton})\,\ln^2 \Lb g^2_0\, \,\frac{1}{Q^2 \,x}\,\Rb
\Rb \right\} \nonumber \eea

One can see, that $F_2\,\, \propto\,\,\alpha'\,Q^2 \,R^2_A\,\Lb \ln
\ln  \Lb g^2_0\, \,\frac{1}{Q^2 \,x}\,\Rb \Rb^2$ .  However, such a
behavior is valid only  in the restricted kinematic region when $
(1/m_{graviton})\,\ln \Lb g^2_0\, \,\frac{1}{Q^2
\,x}\Rb\,\,<\,\,R_A$. For higher energies, we loose all the
specifications related to the nucleus, and the nucleus interacts as
a proton would do, but with the coupling constant $g^2_0\,N_c\,A$.

\section{DIS with nuclei: dipole model.}

 In QCD, the  DIS cross section can be  written as a product of two
factors \cite{DISDIP,MUCD,K}, namely the probability to find a
dipole in the virtual photon, and the scattering amplitude of the
dipole with the target.  In this way the DIS cross section is given
by the expression

\beq \label{DP1}
\sigma_{tot}\Lb DIS;Q^2,x\Rb\,\,=\,\,\int\,\frac{d^2 r\,d \zeta}{2 \pi}\,d^2 b \,|\Psi\Lb Q;r,\zeta\Rb|^2\,N\Lb r,b,x\Rb
\eeq
where $N$ is the imaginary part of the scattering amplitude of the
dipole with size $r$ off the target, and $\zeta$ is the fraction of
the energy carried by the quark of the dipole. We  can try to
generalize this equation to N=4 SYM, namely,

\beq \label{DP2}
\sigma^A_{tot}\Lb DIS;Q^2,x\Rb\,\,=\,\,\int\,\frac{d^2 r\,d \zeta\,d z }{2 \pi}\,d^2 b \,|\Psi\Lb Q;r, z, \zeta\Rb|^2\,N_A\Lb z,b,x\Rb
\eeq

The factorization of \eq{DP1} is valid on very general grounds (see
Ref. \cite{GRIBGL}), and should hold in any reasonable theory, since
it is based on  the structure of the interaction in time. In
\eq{DP2}, we use the fact that the interaction due to the graviton
exchange does not depend on the size of the interacting particles,
(see \eq{N41}).  We do not see any specific features for the dipole
- target interaction, and thus we should be able to use for the
nucleus amplitude ($N_A$) the formulae that we have discussed in the
previous sections.  On the boundary, $\Psi\Lb Q;r,\zeta\Rb$ is known
and it is proportional to $K_1\Lb \bar{Q}\,r\Rb$, or to $ K_0\Lb
\bar{Q}\,r\Rb$,  for different polarizations of the virtual photon
with $\bar{Q}^2 = Q^2 \zeta(1 - \zeta)$. We can reconstruct $\Psi\Lb
Q;r,\zeta\Rb$ using the Witten formula \cite{WIT}, namely,

\bea \label{DP3}
&& \Psi\Lb Q;r,\zeta\Rb\,\,\,=\\
&&\frac{\Gamma\Lb \Delta\Rb}{\pi\,\Gamma\Lb \Delta - 1\Rb}\,\,\int \,d^2 r' \,
\Lb \frac{z}{z^2\,\,+\,\,( \vec{r}\,-\,\vec{r}')^2}\Rb^{\Delta}\,\, K_0\Lb \bar{Q}\,r'\Rb
\,\,\,\mbox{with}\,\,\,\,\Delta_{\pm}\,\,=\,\,\h \Lb d\,\,\pm\,\,\sqrt{d^2 + 4\, m^2}\Rb \nonumber
\eea

Using the formulae {\bf 3.198},  {\bf 6.532(4)}, {\bf 6.565(4)}  and {\bf 6.566(2)} from the Gradstein and Ryzhik Tables, Ref. \cite{RY}, and using the Feynman parameter ($t$),  we can rewrite \eq{DP3} in the form
\bea \label{DP4}
&& \Psi\Lb Q;r,\zeta\Rb\,\,\,=\,\,\frac{\Gamma\Lb \Delta\Rb}{\pi\,\Gamma\Lb \Delta - 1\Rb}\,\int\,\xi\,d \xi\,d^2\, r' \frac{J_0\Lb \bar{Q}\,\xi \Rb}{\xi^2\,+\,r'^2}\,\Lb \frac{z}{z^2\,\,+\,\,( \vec{r}\,-\,\vec{r}')^2}\Rb^{\Delta}\,\,=\\
&&\frac{\Gamma\Lb \Delta + 1\Rb}{\pi\,\Gamma\Lb \Delta - 1\Rb}\,\int \xi\,d \xi\,d^2\, r'\int^1_0\,\frac{d t}{z}\,  t^{\Delta -1}\,(1 - t)\,\,J_0\Lb \bar{Q}\,\xi \Rb\,\Lb \frac{z}{t\,z^2\,\,+\,\,t\,( \vec{r}\,-\,\vec{r}')^2\,\,+\,\,(1 - t)\,r'^2\,+\,(1 - t)\,\xi^2}\Rb^{\Delta + 1}\nonumber\\
&& =\,\,\frac{\Gamma\Lb \Delta\ + 1\Rb}{\pi\,\Delta\,\Gamma\Lb \Delta - 1\Rb}\,\int \xi'\,d \xi'\,\int^1_0\,d t\,  t^{\Delta -1}\,\,J_0\Lb \bar{Q}\,k \Rb\,\Lb \frac{z}{t\,z^2\,\,+\,r^2\,t\,(1 - t)\,\,+\,\,\xi'^2}\Rb^{\Delta }\nonumber\\
&& = \,\,\frac{1}{\pi\,2^{\Delta - 1}\,\Gamma\Lb \Delta - 1\Rb}\,\,z^\Delta \,\int^1_0\,d t\, \Lb \frac{\bar{Q}^2}{\,z^2\,\,+\,(1 - t)\,r^2}\Rb^{\Delta - 1}\,K_{\Delta -1}\Lb \bar{Q}\sqrt{t\,(z^2\,\,+\,(1 - t)\,r^2)} \Rb\nonumber
\eea

Using \eq{DP4}, we can rewrite \eq{DP1} in the form

\bea \label{DP5}
\sigma_{tot}\Lb DIS;Q^2,x\Rb\,\,&=&\,\,\int\,\frac{d^2 r\,d \zeta}{2 \pi}\,d^2 b\,\,d z\, \,\left\{
\,\frac{1}{\pi\,2^{\Delta -1}\,\Gamma\Lb \Delta - 1\Rb}\,\,z^\Delta \,\int^1_0\,d t\, \Lb \frac{\bar{Q}^2}{\,z^2\,\,+\,(1 - t)\,r^2}\Rb^{\Delta - 1}\,\right.\nonumber\\ &\times & \left.
K_{\Delta -1}\Lb \bar{Q}\sqrt{t\,(z^2\,\,+\,(1 -t)\,r^2)} \Rb\,\right\}^2\,
Re\Lb 1\,\,-\,\,\exp\Lb  i\,\frac{ g^2_0\,N_c}{4}\,\,\frac{Q^2}{x}\,z^4\,S\Lb b\Rb\Rb \Rb
\eea

where we used the simple exchange of the graviton as in \eq{DIS4}. Using the asymptotical expression for the modified Bessel  function, we can do the integral over $z$ in saddle point approximation, and the equation for the saddle point $z_{SP}$ has the following form

\beq \label{DP6}
\sum^2_{i=1}\frac{\bar{Q}\,t_i}{\sqrt{t_i(z^2_{SP} \,+\,r^2)}}\,z_{SP}\,\,+\,\,\,i  g^2_0 N_c
\,\frac{Q^2}{x}\,z^3_{SP}\,S\Lb b \Rb\,\,=\,\,0
\eeq
which leads to

\beq \label{DP7}
z_{SP}\,\,=\,\,\sqrt{\frac{ \sum^2_{i=1}\frac{i\,\bar{Q}\,t_i}{\sqrt{t_i\,r^2}}}{\,g^2_0 N_c \frac{Q^2}{x}\,S\Lb b \Rb}}\,\,\propto\,\,\sqrt{\frac{x}{Q\,S\Lb b \Rb \,r}}\,\,\ll\,\,r
\eeq

In \eq{DP5},\eq{DP6} and \eq{DP7},  we  introduced two variables,
$t_1$ and $t_2$, to describe $|\Psi\Lb Q;r,\zeta\Rb|^2$. From
\eq{DP7} and \eq{DP5}, we obtain

\bea \label{DP8}
\sigma_{tot}\Lb DIS;Q^2,x\Rb\,\,&=&\,\,
\int\,\frac{d^2 r\,d \zeta}{2 \pi}\,d^2 b \,\,d \Lb z - z_{SP}\Rb \int^1_0 \prod^2_{i=1} \,\frac{ d t_i}{\sqrt{t_i}}\,\,
\,\frac{\pi\,\Gamma^2\Lb \Delta\Rb}{\pi^2\,\Gamma^2\Lb \Delta - 1\Rb}\,\,\,\Lb \frac{z_{SP}\,\bar{Q}^2}{\,r^2}\Rb^{2\,\Delta - 2}\,\frac{r^3}{\bar{Q}}\\
& &\exp \Lb -\,\sum^2_{i=1}\,\bar{Q}\,\sqrt{t_i\,r^2}\Rb\,\,\left\{\,\,1\,\,-\,\,\exp\Lb  i\,\frac{ g^2_0\,N_c}{4}\,\,\frac{Q^2}{x}\,S\Lb b \Rb\,
\Lb\frac{i\, \sum^2_{i=1}\frac{\bar{Q}\,t_i}{\sqrt{t_i\,r^2}}}{g^2_0 N_c \frac{Q^2}{x}\,S\Lb b \Rb}\Rb^2\Rb
\right\} \nonumber
\eea

Introducing a new variable $\tilde{Q}\,=\,\bar{Q}\,\sum_{i=1}^2\,\sqrt{t_i}$, we can integrate over $r$ using the steepest decent method. The equation for the saddle point reads as follows

\beq
\label{DP9}
r_{SP}\,\,=\,\,\Lb \frac{i\, x}{2\,g^2_0 N_c\,S\Lb b \Rb\,\tilde{Q}}\,\frac{\tilde{Q}^2}{Q^2}\Rb^{\frac{1}{3}}
\eeq

and

\bea \label{DP10}
&&\mbox{the second term in \eq{DP8}}\,\,=\,\,-\,
\int\,\frac{r_{SP} \,d \Lb r - r_{SP}\Rb\,d \zeta}{2 \pi}\,d^2 b \,\,d \Lb z - z_{SP}\Rb \int^1_0  \prod^2_{i=1} \,\frac{ d t_i}{\sqrt{t_i}}\,\, \frac{\pi\,\Gamma^2\Lb \Delta\Rb}{\pi^2\,\Gamma^2\Lb \Delta - 1\Rb}\,\,\,\nonumber \\
&&\,\,\,\,\,\,\,\,\,\,\,\,\,\,\,\,\,\,\,\,\,\,\,\,\,\,\,\,\times
\Lb \frac{z_{SP}\,\bar{Q}^2}{\,r^2_{SP}}\Rb^{2\,\Delta - 2}\,\frac{r^3_{SP}}{\bar{Q}}\,\,
\exp \Lb -\,\Lb - i\,\frac{ Q^2}{Q^2_s}\Rb^{\frac{1}{3}}\,\,+\,\,i \frac{3}{2} \,g^2_0 N_c \frac{Q^2}{x}\,S\Lb b \Rb \frac{1}{r^4_{SP}}\,\,\Lb r - r_{SP}\Rb^2\Rb
\eea

with

\beq \label{DP11}
Q^2_s \,\,=\,\,\Lb \frac{2 \,g^2_0 N_c\,S\Lb b \Rb}{x\,\zeta^4(1 - \zeta)^4\Lb \sum^2_{i=1}\sqrt{t_i}\Rb^4} \Rb\,\,\propto\,
\frac{A^{\frac{1}{3}}}{N_c}\,\frac{1}{x}
\eeq

\eq{DP10} shows geometrical scaling behavior, at least to within
exponential accuracy. The saturation momentum of \eq{DP11},  has
expected from the high density QCD $A$ dependence, increases in the
region of low $x$ in the same way as for the DIS  case, with the
${\cal R}$ current given by \eq{QS}. In general, \eq{DP10} displays
the same features as \eq{DIS5}, (see also  \eq{DIS52}).

\section{DIS in a shock wave approximation.}
The approach developed above, has to be compared with
Ref.\cite{DISKOV}, in which DIS with a nucleus target was considered
in the framework of the shock wave approximation.  In this paper the
usual decomposition  of the DIS cross section into two factors given
by \eq{DP1}\cite{DISDIP,MUCD,K}, which are the probability to find a
dipole in the virtual photon, and the amplitude of the scattering of
the dipole with the target, is  used (see \eq{DP1}) where $\zeta$ is
the fraction of energy carried by the  quark of the dipole. In Ref.
\cite{DISKOV} it is suggested to study the dipole-target amplitude
in the semiclassical limit of the dipole scattering, in the presence
of the shock wave that was produced by the nucleus, in the spirit of
Ref. \cite{HOOF}.  In this approach, the dipole is located at the
boundary of the $AdS_5$ space, and the two-dimensional surface of
the string is characterized by  $X^\mu = X^\mu(\tau, \sigma)$, (with
$\mu = 0,\dots, 4$), which depends on two coordinates $(\tau
,\sigma)$. The string Nambu-Goto action takes the following form

\beq \label{SW2}
S_{NG}\,\,=\,\,\int d \sigma d \tau {\cal L}\,\,=\,\,\frac{1}{ 2 \pi \alpha'}\,\int d \tau d \sigma \sqrt{ - det G_{\alpha, \beta}} \,\,\,\mbox{where}\,\,\,\, G_{\alpha, \beta}\,=\,g_{\mu, \nu}\Lb X\Rb\,\partial_\alpha X^\mu\,\partial_\beta X^\nu, \,\,\,\,\alpha, \beta \,=\,\sigma, \tau
\eeq

In the presence of the heavy nucleus, the free metric of \eq{N44} has to be altered
in order to take into account the energy-momentum tensor that describes the interaction of the dipole string with the nucleus.
 The modified metric is given by

\beq \label{SW3}
d s^2 \,\,=\,\,\frac{L^2}{z^2}\Lb - 2 d x_{+} d x_{-} \,+\,(d x_{\perp})^2 \,+\,d z^2\Rb \,\,+\,\,T_{- -} \delta\Lb x_{-}\Rb d x_- dx_-
\eeq

In \eq{SW3}, we denote $x_{\pm} \,\,=\,\,\frac{x_0 \pm
x_3}{\sqrt{2}}$, where $x_0$ is the time in the normal four
dimensional space. $x_4 \equiv z$. In Ref. \cite{DISKOV}, $T_{-
-}\,\,\mu z^2 \delta( x_)$ , suggested in Ref.\cite{JP},  is used.
Using this assumption of Ref. \cite{DISKOV}, the metric reduces  to
the expression

\bea \label{SW5}
d s^2  &=& \frac{L^2}{z^2}\,\Lb 2 d x_{+}\,d x_{-}\,\,+\,\,\mu\,z^4\,\delta\Lb x_{-}\Rb \,d x_{-}^2\,+\,(d x_{\perp})^2 \,+\,d z^2 \Rb \nonumber \\
d s^2  &=& \frac{L^2}{z^2}\,\Lb 2 d x_{+}\,d x_{-}\,\,+\,\,\frac{\mu}{a}\Theta\Lb x_-\Rb \,\Theta\Lb a - x_-\Rb\,z^4\, \,d x_{-}^2\,+\,(d x_{\perp})^2 \,+\,d z^2 \Rb\nonumber\\
d s^2  &=& \frac{L^2}{z^2}\,\left\{ - \Lb 1 - \frac{\mu}{2 a} z^4 \Rb \,dt^2 \,+\,
\Lb 1 +  \frac{\mu}{2 a} z^4 \Rb \, (d x_3)^2  \, \,+\,\,(d x_{\perp})^2 \,+\,d z^2 \right\}
\eea
where $a$ is chosen such that $\mu/2 a =s^2$, and $a \sim 2
R_A\Lambda/p_+ \propto A^{1/3}/p_+$ (see Ref. \cite{DISKOV} for
details). In the last line of this equation, we omit the theta
functions, since we are looking for the solution which does not
depend on time (static solution \cite{DISKOV}).  The static approximation is not well  justified (see
Ref.\cite{MSX}, which appeared after the first version of this paper we put  on the net). However, the 
exchange of gravito, which interacts with the energy-momentum tensor (see \eq{N41}) and which is responsible for the mediation of the gravitational force, is  taken into account in this approximation. As  we
mentioned above,
 the main goal of this section is to confront the Glauber-Gribov approach for dipole -nucleus scattering,  described in the previous section, with the
static solution in the SW approximation. Although at first glance the solution of Ref. \cite{DISKOV}
does not reproduce the result of the Glauber-Gribov approach (see below),
we will argue, that by changing the form of the Lagrangian of the string
interaction with the nucleus, we are able to reproduce the Glauber-Gribov
formula,
in the static solution. Therefore, although it is plausible that one can
learn some physics from the static solution, we believe however, that we can learn no more than is already derived
from the Glauber-Gribov approach.

  Using \eq{SW2},
\eq{SW5} and the following  parameterization of $X^\mu$, namely $X^0
= t$, $X^1 = x$, $X^2 =0$, $X^3=0$ and $X^4 = z(x)$ as in ref.
\cite{DISKOV}, the action $S$ is found to be equal to

\beq \label{SW51}
S\,\,=\,\,\int^{a \sqrt{2}}_0\,d t \,\int^{r/2}_{- r/2}\,d x \,\,{\cal L}^{static} \,\,\,\,\,\mbox{with}\,\,\,\,\,\,\,
{\cal L}^{static}\,\,=\,\,\frac{\sqrt{\lambda}}{2 \pi}\,\frac{1}{z^2}\,\sqrt{\Lb 1 + z'^2\Rb\,\Lb 1 - s^2\,z^4\Rb}
\eeq

>From the Euler-Lagrange equation, which has the form (for the static solution)

\beq \label{SW6}
\frac{\partial}{\partial x}\,\frac{\partial\,{\cal L}^{static}}{\partial z'}\,\,-\,\,\frac{\partial\,{\cal L}^{static}}{\partial z}\,\,=\,\,0,
\eeq
as in ref. \cite{DISKOV}, the following solution is found

\beq \label{SW7}
S\Lb \mu\Rb\,\,=\,\,\frac{\sqrt{\lambda}\,a}{\pi\,c_0 \sqrt{2}}\,\left\{ \frac{c^2_0\,r^2}{z^3_{max}}\,\,-\,\, \frac{2}{z_{max}}\,\,+\,\,\frac{2}{z_h}\right\}\,\,\,\, \,\,\,\,\,\mbox{with}\,\,\,\,\,\,\,c_0\,=\,\frac{\Gamma^2\Lb 1/4\Rb}{(2 \pi)^{3/2}}\,\,\,\mbox{and}\,\,\,z_h = \frac{1}{\sqrt{s}}
\eeq
while $z_{max} $ is the solution to the equation

\beq \label{SW8}
c_0\,r\,\,\,=\,\,z_{max}\,\sqrt{1\,\,-\,\,s^2\,z^4_{max}}
\eeq

The amplitude $N$ in \eq{DP1} is equal to \cite{DISKOV}

\beq \label{SW9}
N\Lb r, x\Rb \,\,\,=\,\,Re\left\{ 1 - \exp\Lb i S(\mu) \Rb\right\}
\eeq

Using \eq{DP1}, the cross section for DIS has the form

\beq \label{SW10}
\sigma_{tot}\Lb DIS;Q^2,x\Rb\,\,=\,\,\int\,\frac{d^2 r\,d \zeta}{2 \pi}\,|\Psi\Lb Q;r,\zeta\Rb|^2\,N\Lb r,x\Rb
\,\,\propto\,\,\int\,\frac{d^2 r\,d \zeta}{2 \pi}\,K^2_0\Lb \bar{Q} r\Rb\,N\Lb r,x\Rb
\eeq

In \eq{SW10}, we omitted the integration over impact parameter,
since in this simplified string approach we consider that the
nucleus has the infinite extension in the transverse plane. As we
have discussed above, such a simplified approach to the impact
parameter dependence could cause a lot of difficulties, since  DIS
cross sections depend on the impact parameter distribution both in
the nucleus and in the nucleon amplitude (see section 5). In
\eq{SW10} we simplified the wave function of the photon,  which is
known, by replacing it by $K_0$, since at large values of $\bar{Q}^2
= Q^2 \zeta ( 1 - \zeta)$, both components of the photon  wave
function for transverse and longitudinal polarized photons have the
same behavior $\exp\Lb - 2\,\bar{Q}\,r\Rb$.

We expect that in DIS the typical $r$ will be small, and therefore
we try to find the solution to \eq{SW8} for which
$m\,\,\equiv\,\,c^4_0\,r^4\,s^2 \,\ll\,1$.  In this case, in
ref.\cite{DISKOV} three solutions have been found, which correspond
to three different Riemann sheets of the cubic root, and which can
be characterized by the index $n = 0,1,2$. They are

\beq \label{SW11}
z_{max}\,\,\, \xrightarrow{m \to 0}\,\,\, \,=\,\,\,\,\left\{\begin{array}{c} \,\,\,1/\sqrt{s}\,\,\,\,\,\mbox{for}\,\,\,\,\,n=0;\\
\\
\,\,\,i/\sqrt{s}\,\,\,\,\,\mbox{for}\,\,\,\,\,n=1;\\
\\
\,\,\,c_0\,r\,\,\,\,\,\mbox{for}\,\,\,\,\,n=2;\\
\end{array} \right.
\eeq

The solution with $n=2$ is the only one that matches the Maldacena result\cite{MADIS}, for which $m \to 0$.  For this solution, we can take the integral over $r$ in the second term of \eq{SW9} in  \eq{SW10}
by the steepest decent method, with the saddle point

\beq \label{SW12}
r_{SP} \,\,=\,\,\sqrt{ - i \frac{\sqrt{\lambda}\,a}{2\sqrt{2} \pi \bar{Q}}}
\eeq

One can see that

\beq \label{SW13}
m_{SP}\,\,=\,\,c^4_0\,r^4_{SP} \,s^2\,\,=\,\,\frac{ c^4_0 \,\lambda^2\,a^2}{8 \pi^2 \bar{Q}^2} \propto\,
\frac{\lambda^2\,A^{2/3}}{\bar{Q}^2}\,\,\,\ll\,\,1\,\,\,\mbox{for}\,\,\,\bar{Q}^2\,\,\gg\,\,
\lambda^2\,A^{2/3}
\eeq
and, therefore, in the kinematic region $\bar{Q}^2\,\,\gg\,\,
\lambda^2\,A^{2/3}$,  the second term of \eq{SW9} leads to an approach of the unitarity bound for the DIS cross section of the following form

\bea
\sigma_{tot}\Lb DIS;Q^2,x\Rb\,\,&=&\,\,\int\,\frac{d^2 r\,d \zeta}{2 \pi}\,|\Psi\Lb Q;r,\zeta\Rb|^2\,\,\,-
\,\,\sigma_{II} \label{SW141}\\
\sigma_{II}\,\,&=&\,\,\,
\exp\left\{ - \Lb \frac{Q_s}{Q} \Rb^{\frac{1}{2}} \right\} \label{SW142}
\eea

where the pre-exponential factor can be easily calculated.
The saturation momentum $Q_s$ has the form

\beq \label{SW15}
Q_s\,\,=\,\,\frac{\sqrt{\lambda} a\,Q^2 \zeta (1 - \zeta)}{2 \sqrt{2}\,\pi}\,\,\propto \,\,\sqrt{\lambda} A^{1/3}
\,x
\eeq

At large values of $Q$, \eq{SW15} leads to a term of the order of $x A^{1/3}\lambda/Q_s$,
 which corresponds to the twist expansion, with the anomalous dimension $\gamma = 1/2$. The $A$ dependence is in accordance with this as well \cite{LETU}, but the $x$ dependence looks strange. $Q_s \to 0$ at $x \to 0$, and therefore the theory predicts that for low $x$ and $\bar{Q}^2\,\,\gg\,\,
\lambda^2\,A^{2/3}$, the DIS cross section is very small.

It turns out that in the kinematic region $\bar{Q}^2\,\,\ll\,\,
\lambda^2\,A^{2/3}$, the  $n=0$ solutions gives  the largest contribution. Indeed, inserting this solution in \eq{SW7}, one can find the saddle point in  the integration over $r$, which is equal to

\beq \label{SW16}
r_{SP} \,\,=\,\,-i \frac{\pi \sqrt{2}\bar{Q}}{c_0\,\sqrt{\lambda}\,a\,s^{3/2}}\,\,\propto\,i \frac{\bar{Q}}{ \sqrt{\lambda}\,A^{1/3}\,\sqrt{s}}
\eeq

Evaluating $m = c^4_0\,r^2_{SP} \,s^2$, namely

\beq \label{SW17}
m\,\,=\,\,\frac{4 \pi^4 \bar{Q}^4}{\lambda^2\,A^{4/3}}\,\,\ll\,\,1
\eeq
one can see that in the region where $\bar{Q}^2\,\,\ll\,\,
\lambda^2\,A^{2/3}$, we are dealing with small values of $m$, and we
can use the solution of \eq{SW11}. Then $\sigma_{II}$ in this case
is proportional to

\beq \label{SW18}
\sigma_{II}\,\,\,\propto\,\,\exp\Lb - i \frac{Q}{Q_s} \Rb\,\,\,\,
\mbox{with}\,\,Q_s\,\,=\,\,4\frac{c_0\,\sqrt{\lambda}\,a\,s^{3/2}}{\pi \sqrt{2}\zeta (1 - \zeta)}\,\,\propto\,\,\frac{A^{1/3}}{x}
\eeq

The saturation momentum in \eq{SW18} displays all the typical properties that we expect from high density QCD.

It is worthwhile mentioning, that the solution with $n=1$ leads to
$\sigma_{II}\,\,\,\propto\,\,\exp\Lb  \frac{Q}{Q_s} \Rb$, with the same saturation momentum, and it can be selected out since $\sigma$ should be positive.

Both \eq{SW142} and \eq{SW18} have in common the fact that $ z^4_{max}s^2$ turns out to be much smaller than unity ( $ z^4_{max}s^2\,\,\ll\,\,1$). It means that in the general equation for the action of \eq{SW51}, we can consider $z^4 s^2$ to be small, and we expand the action with respect to this parameter. In this case the
contribution at high energy can be reduced to the following action

\beq \label{SW19}
S^{eikonal}\,\,\,=\,\,\frac{\sqrt{\lambda}\,a \,s^2}{\sqrt{2} \pi}\,\int^{r/2}_{-r/2}\,d x\,\, z^2\,\sqrt{1 \,-\,z'^2}
\,\,=\,\,Const\,\, s\,A^{1/3}\int^{r/2}_{-r/2}\,d x\,\, z^2\,\sqrt{1 \,+\,z'^2}
\eeq

This action is closely related to the eikonal formula, as one can see from the second term of \eq{SW19}. Solving the Euler-Lagrange equation of \eq{SW6}, we find that

\beq \label{SW20}
1 + z'^2 = \frac{z^4_{max}}{z^4}
\eeq

which leads to

\beq \label{SW21}
z_{max} = i \frac{\Gamma(1/4)}{\sqrt{\pi}\,\Gamma(3/4)}\,(r/2)
\eeq

Evaluating the integration over $x$ in \eq{SW19}, we obtain the scattering amplitude in the form

\bea \label{SW22}
N\Lb r,s\Rb\,\,&=&\,\,Re\,\left\{ 1 \,\,-\,\,\exp\Lb - Const\,A^{1/3}\,\Lb\frac{\Gamma(1/4)}{\sqrt{\pi} \Gamma(3/4)}\Rb^2\,s\,\Lb \frac{r}{2}\Rb^3 \Rb  \right\}\nonumber \\
 \,\,&=&\,\,Re\,\left\{ 1 \,\,-\,\,\exp\Lb - \kappa\,A^{1/3}\,s\,r^3 \Rb  \right\}
\eea

where we have absorbed all constant factors in the factor $\kappa$. It is easy to see that \eq{SW22} leads to

\beq \label{SW23}
\sigma_{II}\,\,\propto \,\,\exp\Lb - i \Lb \frac{Q}{Q_s}\Rb^{\frac{1}{2}}\Rb
\eeq
with $Q_s$ given by \eq{SW18}. The difference between \eq{SW23} and
\eq{SW18}, as well as the fact that \eq{SW142} does not hold,
requires explanation. Referring back to \eq{SW19},  one can see that
implicitly in $S^{eikonal}$, we neglected the part of the action of
\eq{SW51} which does not depend on $s$. Since this contribution
contains a factor of $a \propto 1/s$ in front, we can expect that
this contribution is negligible at high energy. However, the
integral over $x$ can be divergent and compensates this smallness.
In Ref. \cite{DISKOV}, it was suggested that a subtraction in the
action would cancel the divergence at $z \to 0$.  The eikonal
formula suggests a different type of remedy for this divergence,
namely to introduce the action in the following way (compare with
\eq{SW51})

\bea \label{SW24}
&&S\,\,=\,\,\\
&&\int^{\infty}_{-\infty}\,d t \,\int^{r/2}_{- r/2}\,d x \,\,\Delta {\cal L}^{static}\,\,=\,\,
\int^{a \sqrt{2}}_0\,d t \,\int^{r/2}_{- r/2}\,d x \,\,\Delta {\cal L}^{static} \,\,\,\,\,\mbox{with}\,\,\,\,\,\,\,
\Delta{\cal L}^{static}\,\,=\,\,{\cal L}\Lb T_{\mu \nu}\Rb \,\,-\,\,{\cal L}\Lb T_{\mu \nu}=0\Rb \nonumber
\eea

which leads to

\beq \label{SW25}
\Delta{\cal L}^{static}\,\,=\,\,\frac{\sqrt{\lambda}}{2 \pi}\,\frac{1}{z^2}\,\sqrt{\Lb 1 + z'^2\Rb}\left\{ \sqrt{\Lb 1 - s^2\,z^4\Rb}\,\,-\,\,1\right\}
\eeq

\eq{SW24} has a simple meaning, which is that we need to subtract  the term which is responsible for the movement of the string in empty space
during the period of time of the interaction,
from the interaction induced by the energy-momentum tensor of the nucleus.
One can see that the solution with the action given by \eq{SW24} and \eq{SW25}, reproduces \eq{SW23} (see appendix) .

Therefore, we can conclude that the shock wave approximation can be
reproduced by the  eikonal  formula. It should be stressed that the
eikonal  formula is more general, since in the framework of this
approach we are able to introduce the impact parameter dependence as
well as the quantum corrections related to the reggeized graviton
(Pomeron, see section 5). It is worthwhile mentioning that in
\eq{SW18} the shock wave approximation leads to the same amplitude
as \eq{DP10}, in the dipole approach. However, it should be stressed
that the main result of Ref.\cite{DISKOV}, that the dipole amplitude
at high energy has a form
\beq  \label{KR} N(r) \,\,\propto \,\,1 -
\exp^{-\,r Q_s}\,\,\mbox{with}  \,\,Q_s\,\, \propto\,\,Const(x)
A^{1/3} \eeq
holds  in the approach with the action given by
\eq{SW24}. Indeed, this result does not depend on the modification
of the Lagrangian since for $z^4S^3 \,\,\gg\,\,1$ the action is the
same in both approaches  , namely,
\beq \label{SW26} S\,\,=\,\,i
\int^{a \sqrt{2}}_0\,d t \,\int^{r/2}_{- r/2}\,d x
\,\,\frac{\sqrt{\lambda}}{2 \pi}\,s\,\,\sqrt{1  + z'^2}\,\,\to\,\,
i\, Const(x)\, A^{1/3} \,r\, \eeq
The last equation comes from the
equation of motion which leads to $z'=0$.  This saturation momentum
$Q_s(A) \propto Const(x) A^{1/3}$  needs an explanation since it
does not appear in the  Glauber-Gribov approach.  First, the
contribution of \eq{KR} to the DIS cross section (see \eq{DP1}) has
the form
\beq \label{SW27} \sigma\Lb DIS; Q^2,
x\Rb\,\,=\,\,\,\frac{\pi}{Q^2}\,\,\left\{   1 \,+\,\frac{
\pi^2}{16}\,\frac{Q_s(A)}{Q}\,{}_2F_1\Lb \frac{3}{2},\frac{3}{2},2,
\frac{Q^2_s(A)}{4\,Q^2}\Rb -
{}_3F_2\Lb\{1,1,1\},\{\frac{1}{2},\frac{3}{2}\},\frac{Q^2_s(A)}{4\,Q^2}\Rb\right\}
\eeq
For $Q\,>\, Q_s(A)$ $ \sigma\Lb DIS; Q^2, x\Rb\,\to\, \pi/Q^2$
which corresponds to $1$ in \eq{KR}. If we replace $K_0$ by its
asymptotic behavior $ \sigma\,\propto \,(1 /Q^2)\times(1/(Q +
Q_s(A))$.  The physical meaning of $Q_s(A)$ is rather obvious:
during the passage of the dipole through the nucleus the transverse
momentum ($Q$) can get an additional momentum $\Delta Q$ due to
elastic  rescattering with the nucleons , namely
\beq \label{SW28}
\Delta Q\,\,\propto\,q^N_{\perp}\,\times\,\mbox{number of
collisions}\,\,=\,\,\frac{1}{R_N}\,A^{1/3} \eeq
where $q^N_{\perp}
\,=\,1/R_N$ is the typical transverse momentum for elastic
scattering with one nucleon ($R_N$ is the nucleon radius). In the
Glauber-Gribov approach, however, $q^N_{\perp} \propto 1/R_A \to 0$
due to the nucleus form factor (see  \eq{GA8}). In the shock wave
approach the nucleus  wave function has not been taken into account
and  nucleons can have unrestricted transverse momenta.  Therefore,
we consider this momentum as the artifact of the shock wave model in
which, we believe, we need to specify the DIS as scattering with $Q
>  \Delta Q \approx Q_s(A)$ if we are interested in finding  the
total cross section. However, this $Q_s(A)$ can manifest itself in
the inclusive production leading to the situation with two
characteristic momenta that has been advocated in Ref. \cite{MU2SC}.
It should be stressed that \eq{SW28} is written for a string with
the fixed transverse coordinate  (see \eq{SW51} $ X^1 = x$,  $X^2 =
0$). In the general case $\Delta Q^2 \,=\,\frac{1}{R_N}\,A^{1/3}$.
The second comment on \eq{SW28} is that we considered the
rescatterings which are instantaneous  in accordance with the static
solution.  In the region $Q > Q_s(A)$, the contribution, given by \eq{SW27}, is small and  the value of the total cross section for DIS is determined by the saddle point approximation (see \eq{SW23}) which is the same both in the shock wave
approximation and in the Glauber-Gribov approach.

\section{Conclusions}

It is our  common wisdom nowadays that N=4 SYM , which can be solved
at large coupling values, can provide us with some knowledge of what
potentially lies in the confinement region of QCD. However, the
first analysis of high energy DIS scattering, performed in Refs.
\cite{BST1,MHI}, demonstrated that the high energy scattering in N=4
SYM looks quite different from what has been known so far. Contrary
to the usual expectations based on perturbative QCD and the parton
model,
 that the main process at high energies is multiparticle production, it was found in Refs.~\cite{BST1,MHI} that in N=4 SYM the major contribution originates from quasi-elastic scattering. This also contradicts what is known from data.

The goal of this paper was to develop the Glauber-Gribov description
of DIS  on a nuclear target within the N=4 SYM, which should help to
see the key features of high energy scattering in a more transparent
way. For this purpose we employed the eikonal approximation which
has been developed for N=4 SYM in Refs.\cite{BST1,BST2,BST3,COCO,MHI}.
Our results can be summarized as follows.

 \begin{enumerate}
 \item
 We derived the Glauber-Gribov  formula (see \eq{GLF5} and \eq{DIS5}),
and showed that for the case of graviton exchange, this formula displays
the same general properties, such as the geometrical scaling behavior, as in the case of the high density QCD approach.\\

\item
We demonstrated that graviton exchange indeed leads to a total cross
section which is dominated by quasi-elastic re-scatterings. However,
we found that the quantum effects responsible for graviton
reggeization give rise to an imaginary part of the nucleon
amplitude. This imaginary part, enhanced by multiple interactions,
results in a DIS which looks similar to one predicted by the high
density QCD,
(see \fig{totin}).\\

\item
We concluded that in N=4 SYM the impact parameter dependence of the amplitude is essential,  and the entire kinematic region can be divided into three regions. In the first region ($z^2\,g^2_0 N_c \leq 1$),  we can use the eikonal formula with a single graviton or reggeized graviton exchange for the nucleon amplitude. In the second kinematic region, $z^2\,g^2_0 N_c \geq 1$ but $ b^2_0 \propto z^2 s < 1/m^2_{graviton}
<R^2_A$, the multi-graviton exchange in the nucleon amplitude may become important. However, we found that this is not the case and still the single graviton exchange dominates. In the third kinematic region ($z^2\,g^2_0 N_c \geq 1$ and $   1/m^2_{graviton}\,<\,b^2_0 \propto z^2 s\,<\,R^2_A$), the multi-graviton exchanges in the nucleon amplitude must be included, and the related modification to the amplitude are discussed in section 5.3.
\end{enumerate}

In this paper, we considered mostly the DIS of the ${\cal R}$
current with the target. However, in the last two sections, we
discussed the traditional  approach to DIS based on the
factorization given by \eq{DP1}. We considered DIS in two different
ways. In the first one we generalized the usual dipole formula to
N=4 SYM. We derived the probability to  find a dipole in the virtual
photon, in $AdS_5$ space, and considered for the dipole scattering
amplitude the eikonal formula. In the second approach, we revisited
the shock wave approximation that has been developed for DIS in Ref.
\cite{DISKOV}, and we showed that in this formalism we can also use
the Glauber-Gribov approach for DIS in the region of $r \approx
Q/A^{1/3}\sqrt{s}$ . However, the Glauber-Gribov approach suggests a
different way to renormalize the interaction Lagrangian proposed in
Ref.\cite{ DISKOV}. After such modification of the original
formalism of Ref. \cite{DISKOV}, both approaches, namely, the dipole
model and the shock wave approximation give the same result for  $r
\approx Q/(A^{1/3}\sqrt{s})$. We gave the interpretation of the
appearance of the new saturation momentum $Q_s(A)$  that does not
depend on energy\cite{DISKOV} and argue that in the shock wave
approximation we should consider only DIS with $Q
> Q_s(A)$. For such large values of $Q$ the shock wave approximation with our modified Lagrangian reproduces the same result as the Glauber-Gribov approach.

In general, we conclude that N=4 SYM does not lead to any obvious contradiction, either with the high density QCD, or with experimental data. Therefore, we hope to learn something valuable about the confinement region from the exact solution in N=4 SYM, relying on the AdS/CFT correspondence.

\section* {Acknowledgements}
One of us (E.L.)  is very  grateful  to   Yura Kovchegov  and
Chung-I Tan  for their exceptional patience in answering questions
on the shock wave approximation and on  N=4 SYM at high energy.  He
thanks them and Lev Lipatov  for beautiful discussions on the
subject. Our special thanks go to Yura Kovchegov for reading the
manuscript and for his comments improving the presentation. E.L.
also thanks the  high energy theory group of the University Federico
Santa Maria for the hospitality and creative atmosphere.

 This work was supported in part by Fondecyt (Chile) grants, numbers 1050589,
7080067 and 7080071, by DFG (Germany)  grant PI182/3-1,
 by BSF grant $\#$ 20004019, by
a grant from Israel Ministry of Science, Culture and Sport and
the Foundation for Basic Research of the Russian Federation.

\appendix

\renewcommand{\theequation}{A-\arabic{equation}}
\setcounter{equation}{0}  

\begin{boldmath}
\section{Shock wave approximation for DIS with our hypohesis on renormalized Lagrangian} \label{sec:A}
\end{boldmath}

In this appendix we consider the shock wave approximation to DIS suggested in Ref. \cite{DISKOV}, with our hypothesis on the renormalised Lagrangian. As has been mentioned, we assume that
the static $AdS_5$ renormalised lagrangian is the regular $AdS_5$ lagrangian with a nucleus present, minus the vacuum $AdS_5$ lagrangian, where the nucleus is not present. The expression to such a renormalised lagrangian is given by the following expression

\bea
\mL^{\mbox{ren}}\,&&={\cal L}\Lb T_{\mu \nu} \Rb \,-\,{\cal L}\Lb T_{\mu \nu} =0\Rb \,\,=\,\,\mL^{\mbox{nuc}}\,-\mL^{\mbox{vac}}\,\label{app1}\\
\nonumber\\
\mbox{where}\,\,\,\,\,\,\,\mL^{\mbox{nuc}}\,&&=\,\f{\sqrt{2\la}}{2\pi}\f{1}{z^2}\sqrt{\Lb 1+z^{\,\prime\,2}\Rb\Lb 1-s^2z^4\Rb}\,\,\,\,\,\,\mbox{and}\,\,\,\,\,\,\mL^{\mbox{vac}}\,=\,\f{\sqrt{2\la}}{2\pi}\f{1}{z^2}\sqrt{\Lb 1+z^{\,\prime\,2}\Rb}
\eea

The Euler - Lagrange equation for $\mL^{\mbox{ren}}$ takes the form;

\bea
\f{\D\mL^{\mbox{ren}}}{\D\,z}\,-\,\f{\D}{\D\,x}\Lb\f{\D\mL^{\mbox{ren}}}{\D\,z^{\,\prime\,}}\Rb\,&&=0\nonumber\\
\Rightarrow\,\,\,\,\,\,\f{\D\mL^{\mbox{nuc}}}{\D\,z}\,-\,\f{\D\mL^{\mbox{vac}}}{\D\,z}\,\,-\,\f{\D}{\D\,x}\Lb\f{\D\mL^{\mbox{nuc}}}{\D\,z^{\,\prime\,}}\Rb\,+\,\f{\D}{\D\,x}\Lb\f{\D\mL^{\mbox{vac}}}{\D\,z^{\,\prime\,}}\Rb&&=0\label{EL}
\eea

The various terms appearing in \eq{EL} can be calculated from \eq{app1}, namely

\bea
\f{\D\mL^{\mbox{nuc}}}{\D z}\,&&=\,-\,\f{2}{z}\f{\mL^{\mbox{nuc}}}{\Lb 1-s^2z^4\Rb}\,\,\,\,\,\,\,\,\,\,\,\,\,\,\,\,\,\,\,\,\,\,\,\,\,\,\,\f{\D\mL^{\mbox{vac}}}{\D z}\,=\,-\,\f{2}{z}\mL^{\mbox{vac}}\label{app2}\\
\nonumber\\
\mbox{and}\,\,\,\,\f{\D}{\D x}\Lb\f{\D \mL^{\mbox{nuc}}}{\D z^{\,\prime}}\Rb\,&&=\,\f{\D}{\D\,x}\Lb\f{z^{\,\prime}}{1+z^{\,\prime\,2}}\,\mL^{\mbox{nuc}}\Rb\nonumber\\
&&=\,\Lb\f{z^{\,\prime\,\prime}}{1+z^{\,\prime\,2}}\,-\f{2z^{\,\prime\,2}z^{\,\prime\,\prime}}{\Lb 1+z^{\,\prime\,2}\Rb^2}\Rb\mL^{\mbox{nuc}}\,+\,\f{z^{\,\prime\,2}}{1+z^{\,\prime\,2}}\f{\D \mL^{\mbox{nuc}}}{\D z}\nonumber\\
&&+\,\f{z^{\,\prime}\,z^{\,\prime\,\prime}}{1+z^{\,\prime\,2}}\f{\D\mL^{\mbox{nuc}}}{\D z^{\,\prime}}\nonumber\\
\nonumber\\
&&=\,\Lb\f{z^{\,\prime\,\prime}}{1+z^{\,\prime\,2}}\,-\f{z^{\,\prime\,2}z^{\,\prime\,\prime}}{\Lb 1+z^{\,\prime\,2}\Rb^2}\Rb\mL^{\mbox{nuc}}\,+\,\f{z^{\,\prime\,2}}{1+z^{\,\prime\,2}}\f{\D \mL^{\mbox{nuc}}}{\D z}\label{app3}\\
\nonumber\\
\mbox{similarly}\,\,\,\,\f{\D}{\D x}\Lb\f{\D\mL^{\mbox{vac}}}{\D z^{\,\prime}}\Rb&&=\,\Lb\f{z^{\,\prime\,\prime}}{1+z^{\,\prime\,2}}\,-\f{z^{\,\prime\,2}z^{\,\prime\,\prime}}{\Lb 1+z^{\,\prime\,2}\Rb^2}\Rb\mL^{\mbox{vac}}\,+\,\f{z^{\,\prime\,2}}{1+z^{\,\prime\,2}}\f{\D \mL^{\mbox{vac}}}{\D z}\label{app4}
\eea

Plugging \eq{app2}, \eq{app3} and \eq{app4} into \eq{EL} gives the result

\bea
\f{1}{1+z^{\,\prime\,2}}\Lb\f{\D\mL^{\mbox{nuc}}}{\D\,z}\,-\,\f{\D\mL^{\mbox{vac}}}{\D\,z}\,\Rb\,-\,\Lb\f{z^{\,\prime\,\prime}}{1+z^{\,\prime\,2}}\,-\f{z^{\,\prime\,2}z^{\,\prime\,\prime}}{\Lb 1+z^{\,\prime\,2}\Rb^2}\Rb\Lb\mL^{\mbox{nuc}}\,-\,\mL^{\mbox{vac}}\Rb&&=0\nonumber\\
\Rightarrow\,\,\,\,\,\,\,\,-\,\f{2}{z}\f{\mL^{\mbox{nuc}}}{\Lb 1-s^2z^4\Rb}\,+\,\f{2}{z}\,\mL^{\mbox{vac}}\,\,-\,\f{z^{\,\prime\,\prime}}{1+z^{\,\prime\,2}}\,\Lb\mL^{\mbox{nuc}}\,-\,\mL^{\mbox{vac}}\Rb&&=0\nonumber\\
\Rightarrow\,\,\,\,\,\,\,\,\,\,\,\,\,\,\,-\,\f{2}{z}\f{\Lb 1-\sqrt{1-s^2z^4}\Rb}{\sqrt{1-s^2z^4}}\,\,+\,\f{z^{\,\prime\,\prime}}{1+z^{\,\prime\,2}}\,\Lb 1-\sqrt{1-s^2z^4}\Rb&&=0\nonumber\\
\Rightarrow\,\,\,\,\,\,\,\,\,\,\,\,\,\,\,\,\,\,\,\,\,\,\,\,\,\,\,\,\,\,\,\,\,\,\,\,\,\,\,\,\,\,\,\,\,\,\,\,\,\,\,\,\,\,\,\,\,\,\,\,\,\,\,\,\,\,2\Lb 1+z^{\,\prime\,2}\Rb\,-\,z\,z^{\,\prime\,\prime\,}\,\sqrt{1-s^2z^4}\,&&=\,0
\label{EL2}
\eea

Recall that one can express $z^{\,\prime\,\prime\,}$ as $\Lb 1/2\Rb\,\D z^{\,\prime\,2}/\D z$, hence \eq{EL2} simplifies to

\bea
2\Lb 1+z^{\,\prime\,2}\Rb\,&&=\,\h\,z\,\f{\D z^{\,\prime\,2}}{\D z}\,\sqrt{1-s^2z^4}\,\,\,\,\,\,\Rightarrow\,\,\,\,\,\,\f{d z^{\,\prime\,2}}{1+z^{\,\prime\,2}}\,=\,\f{dz}{z}\,\f{4}{\sqrt{1-s^2z^4}}
\label{EL3}
\eea

Integrating between $z^{\prime}\Lb z\Rb$ and $z^{\,\prime}\Lb z_m\Rb\,=\,0$, where $z_m$ is an extremum point of the string, one arrives at the result

\bea
z^{\,\prime\,2}\,&&=\,\Lb\f{z}{z_m}\Rb^4\,\Lb\f{1+\sqrt{1-s^2z^4_m}}{1+\sqrt{1-s^2z^4}}\Rb^2\,-1\label{sol}
\eea

>From \eq{sol}, one can find that

\beq \label{AS1}
H\Lb \xi,\xi_m\Rb \equiv\,\int^\xi_0\frac{ d \xi'}{ \sqrt{\Lb\f{\xi'}{\xi_m}\Rb^4\,\Lb\f{1+\sqrt{1 - \xi^4_m}}{1+\sqrt{1-\xi'^4}}\Rb^2\,-1}} \,\,=\,\,
\,\sqrt{s}\,\Lb x \,\,-\,\,r/2\Rb
\eeq

where $\xi = \sqrt{s} z$. In \eq{AS1}, the half of the string where
$z^{\,\prime}\,>\,0$ is chosen, and we integrated over $x$ from
$-r/2$ to $x$. The maximal value of $\xi = \xi_m$, can be found from
the following equation

\beq \label{AS2}
 H\Lb \xi_m,\xi_m\Rb \,\,=\,\,-\sqrt{s}\,r/2
\eeq

We have not yet found the expression for the function  $H\Lb
\xi_m,\xi_m\Rb$  through known functions, but the figure of
\fig{hfun} and \fig{hfunl} demonstrates the behavior of this
function. The key difference with the solution proposed in Ref.
\cite{DISKOV} is the fact that  $H\Lb \xi_m,\xi_m\Rb$ of \eq{AS2}
has only one solution in the region of small $\xi_m$, while $H\Lb
\xi_m,\xi_m\Rb$ of Ref. \cite{DISKOV} has two solutions (see
\fig{hfun}). We can simplify the integrand by its expression at low
$\xi$, namely,

\bea \label{AS3}
H^{\mbox{Low $\xi$}}\Lb \xi,\xi_m\Rb \,\,\to\,\int^{\xi}_0\f{d\xi^{\,\prime}}{\sqrt{\xi^{\,\prime\,4}/\xi^4_m-1}}\,=\,\int^{\xi/\xi_m}_{0}\f{d\zeta\xi_m}{\sqrt{\zeta^4-1}} \eea

Changing the integration variable to $\zeta^2\,=\,\sin\theta$, then \eq{AS3} becomes;

\bea \label{AS3a}
H^{\mbox{Low $\xi$}}\Lb \xi,\xi_m\Rb \,\,\to\,=\,\f{i}{2}\,\xi_m\,\int^{\arcsin\Lb\xi^2/\xi^2_m\Rb}_{0}\f{d\theta}{\sqrt{\sin\theta}} \eea

Finally changing the variable of integration once again to $\sqrt{\sin\theta}\,=\,t$, \eq{AS3a} reduces to

\bea \label{AS3b}
H^{\mbox{Low $\xi$}}\Lb \xi,\xi_m\Rb \,\,\to\,&&=\,i\,\xi_m\,\int^{\xi/\xi_m}_{0}\f{dt}{\sqrt{1-t^2}} \,=\,\mbox{ellpt}\left\{ \arcsin\Lb \xi/\xi_m\Rb\,,0\right\}\notag\\
&&=\,\sqrt{s}\Lb x-r/2\Rb\eea
where $\mbox{ellpt}\Lb\phi,k\Rb$ is the elliptic function defined as

\bea
\mbox{ellpt}\Lb \phi,k\Rb\,=\,\int^{\sin\phi}_0\f{dt}{\sqrt{1-k^2t^2}\sqrt{1-t^2}}\label{AS3c}
\eea

At large values of $\xi$, expanding the integrand at large values of $\xi$, we obtain

\beq \label{AS4}
H^{\mbox{High $\xi$}}\Lb \xi,\xi_m\Rb\,\,\to\,\,i \frac{\xi^2_m}{\sqrt{2 + 2 \sqrt{1 - \xi^4_m}}}\int^\xi_0 \,d \xi' \,\,=\,\,i\,\,\,\frac{\xi^2_m\,\xi}{\sqrt{2 + 2 \sqrt{1 - \xi^4_m}}}\,\,=\,\,
\,\sqrt{s}\,\Lb x \,\,-\,\,r/2\Rb
\eeq

\fig{hfun} and \fig{hfunl} show how the simplified equations (\eq{AS3} and \eq{AS4}), describe the exact function  $H\Lb \xi_m,\xi_m\Rb$ given by \eq{AS1}.

\DOUBLEFIGURE[h]{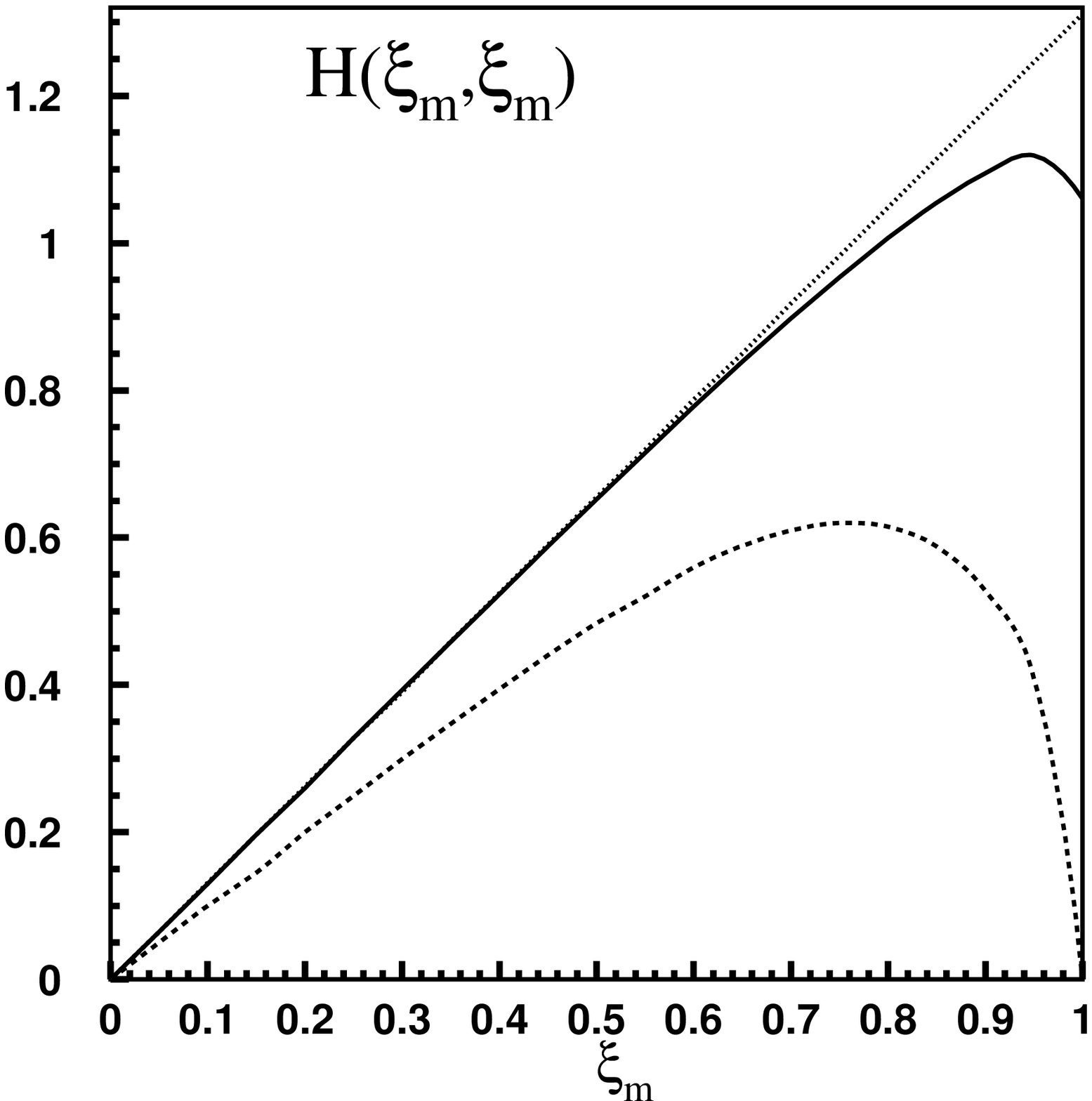,width=85mm,height=75mm}{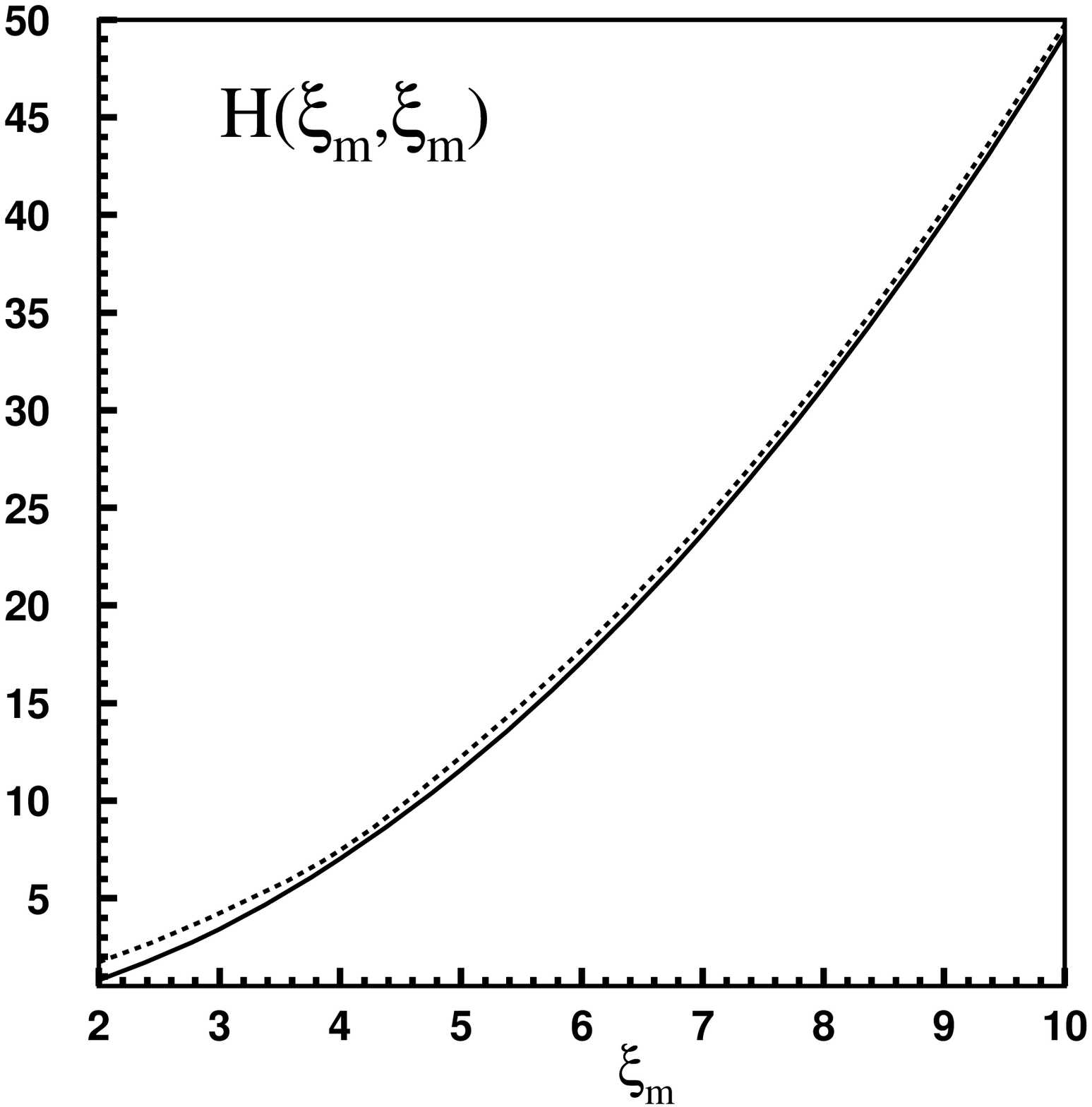,width=85mm,height=75mm}
{Function $H\Lb \xi_m,\xi_m\Rb$ versus $\xi_m$ for small values of $\xi_m$. The solid line shows the function  $H\Lb \xi_m,\xi_m\Rb$
given by \protect\eq{AS1}, and the dotted line is the same function for the solution given in Ref. \protect\cite{DISKOV} while dashed line describes the approximation of \protect\eq{AS3}.
\label{hfun}}
{Function $H\Lb \xi_m,\xi_m\Rb$ versus $\xi_m$ for large values of $\xi_m$. The solid line shows the function  $H\Lb \xi_m,\xi_m\Rb$
given by \protect\eq{AS1}, dotted line is the approximation by \protect\eq{AS4}. \label{hfunl}}

Using \eq{AS3}, one can easily see that \eq{SW51} with  Lagrangian of \eq{app1}  reproduces \eq{SW23},
which we obtain from the eikonal formula.

\end{document}